# Giant Magnetization Enhancement at Ultrathin Limit of an Orthoferrite

*Jin Young Oh[1], Yaolong Xing[2], Seung Gyo Jeong[1], Sehwan Song[3], Woo-suk Noh[4], Valeria Lauter[5], Seongjae Choi[6], Seyoung Kwon[3], Jaeha Choi[3], Si Won Kim[1], Hong Hyun Jeon[1], Heemin Lee[7], Cheng-Tai Kuo[7], Jun-Sik Lee[7], Chanyoung Lee[8,9], Yeonkyu Lee[8,9], Jeehoon Kim[8,9], Heung-Sik Kim[2], Sungkyun Park[3*], Sang Ho Oh[2*], Woo Seok Choi[1*]*

[1]*Department of Physics, Sungkyunkwan University, Suwon 16419, South Korea*

[2]*Department of Energy Engineering, Korea Institute of Energy Technology, Naju 58330, South Korea*

[3]*Department of Physics, Pusan National University, Busan 46241, South Korea*

[4]*MPPC-CPM, Max Planck POSTECH Korea Research Initiative, Pohang 37673, South Korea*

[5]*Spallation Neutron Source, Oak Ridge National Laboratory, Tennessee 37831, USA*

[6]*Department of Semiconductor Physics, Kangwon National University, Kangwon 24341, South Korea*

[7]*Stanford Synchrotron Radiation Lightsource, SLAC National Accelerator Laboratory, California 94025, USA*

[8]*Department of Physics, Pohang University of Science and Technology, Pohang 37673, South Korea*

[9]*Center for Quantum Dynamics of Angular Momentum, Pohang University of Science and Technology, Pohang 37673, South Korea*

J. Y. Oh and Y. Xing contributed equally to this work.

[*]Corresponding authors: psk@pusan.ac.kr, shoh@kentech.ac.kr, choiws@skku.edu

**Abstract**
Low-dimensional magnetic systems have received significant interest due to their high scalability and energy efficiency for future electromagnetic applications. However, low-dimensional magnetism at finite temperature is subject to Mermin–Wagner fluctuations, which make its realization highly challenging. Here, we propose structural modulation in the thin-film structure of an orthoferrite system to achieve a high low-dimensional magnetic response at room temperature. $ErFeO_3$ epitaxial thin film with atomically precise thickness control demonstrates that the magnetic response at 300 K can be significantly enhanced compared to its bulk counterpart. The magnetic response reaches its maximum value of 3.7 $\mu_B$/f.u. for the 2 nm-thick epitaxial film. The characteristic domain structure resulting from epitaxial strain and orbital reconstruction at the surface, both arising from the thin film geometry, has been proposed as the microscopic origin of the giant magnetization enhancement. We show that magnetic ordering can indeed be realized in the 2D limit, with deliberate structural control of orthoferrite systems.

**Keywords**: epitaxial thin film, low-dimensional system, magnetism, orthoferrite, pulsed laser epitaxy

## 1. Introduction

Room temperature magnetization in low-dimensional materials is challenging to achieve. Mermin-Wagner theorem dictates that 2D systems cannot sustain long-range magnetic ordering at a finite temperature [1]. Even with strong anisotropy, thermal fluctuations suppress the realization of high-temperature magnetism in ultrathin systems. Nevertheless, when realized, such systems serve as ideal building blocks for spintronic and memory applications with high scalability [2-4]. Hence, the discovery of suitable magnetic thin films and the development of tuning strategies are essential.

Orthoferrites exhibit a variety of electromagnetic properties stemming from their characteristic orthorhombic lattice. The combination of the orthorhombic crystal structure and spin-orbit coupling introduces Dzyaloshinskii-Moriya interaction (DMI). With additional magnetic contribution from rare earth elements, non-trivial magnetic phases are observed in orthoferrites including $ErFeO_3$, $DyFeO_3$, and $TbFeO_3$ [5-7]. Strong magnetoelectric coupling establishes unconventional ferroelectricity in $HoFeO_3$ and $LuFeO_3$, and an exchange bias effect was observed in the ferroelectric hysteresis of $DyFeO_3$ [8-11]. Uniaxial stress and resultant spin-canting induce multiferroicity in $DyFeO_3$ and $SmFeO_3$, respectively [12,13]. Owing to their complex and peculiar magnetic structures that are strongly coupled to the lattice, orthoferrites have been extensively studied for their electromagnetic functional properties, including doping-induced tunable magnetic interactions at high temperatures and ultrafast spin/magnon dynamics [14-16].

Bulk $ErFeO_3$, an insulating orthoferrite with the magnetic rare earth element Er, has rich magnetic phases due to non-trivial exchange interactions between and among $Er^{3+}$ and $Fe^{3+}$ ions. At 300 K, bulk $ErFeO_3$ exhibits orthorhombic structure with the Pbnm space group and a G-type canted antiferromagnetic phase $\Gamma_4(G_x, A_y, F_z)$ with antiferromagnetic vector $G$ along the $a$-axis and a weak ferromagnetic moment $F$ along the $c$-axis. The canted antiferromagnetic phase with high Néel temperature (~640 K) arises from competition between the $Fe^{3+}$-$Fe^{3+}$ exchange interaction and the DMI interaction between $Fe^{3+}$-$Er^{3+}$ ions. During the spin reorientation transition $\Gamma_{24}(G_{xz}, F_{xz})$ between 97 and 87 K, which is derived from $Fe^{3+}$-$Er^{3+}$ anisotropic symmetric exchange, the weak ferromagnetic moment $F$ rotates from the $c$-axis to the $a$-axis within the $ac$-plane and finally become $\Gamma_2(F_x, C_y, G_z)$. Below 4.5 K, exchange interaction between $Er^{3+}$ ions lead to long-range ordering of $Er^{3+}$ moments [17-19].

Using the controllable tuning knobs offered by thin film geometry, e.g., epitaxial strain, thickness with atomic-scale precision, and spatial dimensionality, the relationship between the electromagnetic properties and the lattice of orthoferrite materials can be elucidated. For example, orthoferrite thin films including $DyFeO_3$, $LuFeO_3$, and $EuFeO_3$ show spin-flop switching, modulation of magnetic phases under epitaxial strain, and magnetic enhancement due to defect engineering [20-22]. In addition, $YFeO_3$ and $LuFeO_3$ thin films show ferroelectricity at 300 K that is derived from antisite defect [23,24]. These results support the idea that thin-film structure provides a promising route to tuning and understanding electromagnetic properties inaccessible in bulk systems. Furthermore, the unconventional characteristics of orthoferrite, e.g., magnetic anisotropy, octahedral tilt, and DMI, can help stabilize the low-dimensional magnetism beyond the Mermin-Wagner limit [25,26]. Such properties are more evident in orthoferrites with heavy rare-earth elements, such as $ErFeO_3$, than others, because the large orbital angular momentum and spin-orbit coupling of localized 4$f$ electrons in $Er^{3+}$ strongly affect $Fe^{3+}$-$Er^{3+}$ exchange interaction, which realizes spin canting of $Fe^{3+}$ moments even at room temperature, despite the absence of ordering of $Er^{3+}$ moments.

Moreover, the interplay of such exchange interaction with structural modulation induced by thin film geometry may give rise to novel magnetic functionalities.

In this study, we discovered giant magnetization enhancement in $ErFeO_3$ epitaxial thin films as their thickness approaches the ultrathin limit. We chose $ErFeO_3$ epitaxial films as a model system to manifest ultrathin magnetization, as we prepared a few-nm-thick samples with atomic-scale precision using pulsed laser epitaxy (PLE). Magnetic characterizations consistently reveal a giant enhancement of magnetization as the film thickness decreased to 2 nm, reaching a maximum saturation magnetization ($M_s$) value of ~3.7 $\mu_B$/f.u. at 300 K. Combined with insights from first-principles calculation, we propose that the giant magnetic response originates from the contributions of octahedral distortion, structural/magnetic domain formation, and exchange interaction variation at the surface and interface in $ErFeO_3$ ultrathin films. The discovery of giant magnetic enhancement at the ultrathin limit, which is counterintuitive to conventional wisdom, would lead to engineering possibilities for low-dimensional magnetic materials.

**2. Results and Discussion**

Orthoferrite $ErFeO_3$ epitaxial thin films are fabricated on cubic $SrTiO_3$ (001) substrates with sharp surfaces and interfaces by PLE down to the ~1.2 nm (~3 pseudocubic unit cells; u.c., 1 u.c. = 0.3796 nm in bulk) thickness. Figure 1a shows X-ray reflectivity of $ErFeO_3$ thin films with 1.2 to 3.0 nm of thickness. The clear thickness fringes persisting up to > 10 $nm^{-1}$ indicate atomically sharp surfaces and interfaces with well-defined global thickness. Figure 1b shows X-ray diffraction (XRD) $\theta$-$2\theta$ scans exhibiting pseudocubic $ErFeO_3$ (00*l*) peaks along with substrate $SrTiO_3$ (00*l*) peaks. Atomic force microscopy images present a clear step-terrace structure, as shown in the inset of Figure 1b. Figure 1c shows high-angle annular dark-field (HAADF) scanning transmission electron microscopy (STEM) and energy dispersive spectroscopy (EDS), manifesting a high-quality epitaxial thin film with atomically sharp interface. Potential cation intermixing at the interface is limited to < 1 u.c., which confirms the quantitative identification of the film thickness. Despite the large lattice mismatch (~2.79%) between $ErFeO_3$ (0.3796 nm) and $SrTiO_3$ (0.3905 nm), the $ErFeO_3$ films with thickness up to ~10 nm exhibit a fully strained nature according to the STEM images. However, XRD reciprocal space map (RSM) of the 20-nm-thick $ErFeO_3$ film shows strain relaxation behavior (Figure S1), as expected.

We confirm exclusive $Fe^{3+}$ valence in the $ErFeO_3$ epitaxial thin films, with signs of spin ordering. Figures 1d,e show X-ray absorption spectroscopy (XAS) results of 2-nm-thick $ErFeO_3$ film for Fe $L_{2,3}$-edges at 300 K, with 45° incident angle and 7 T of magnetic field along the X-ray direction. Note that the 0 T results are more or less identical, as shown in Figure S2. Splitting of Fe $L_3$-edge (~708.5 and 710.2 eV) and $L_2$-edge (~721.6 and 723.3 eV) absorption spectra is consistent with that of the normalized $Fe^{3+}$ $L$-edge XAS of stoichiometric $BiFeO_3$ [27]. The predominant $Fe^{3+}$ valence suggests stoichiometric $ErFeO_3$ without a substantial amount of oxygen vacancies or cation intermixing, which may affect magnetic properties of the system. Horizontal and vertical (Figure 1d) polarization spectra are used to deduce X-ray linear dichroism (XMLD), resulting from conventional antiferromagnetic ordering [28]. Figure 1e shows a clear XMLD signal at the Fe $L_3$-edge for both 0 and 7 T of magnetic field, suggesting overall antiferromagnetic $Fe^{3+}$ ordering, which enhances at low temperature (Figure S2).

An emergent 300 K magnetization of $ErFeO_3$ epitaxial thin film is observed, with a maximized $M_s$ of 3.7 $\mu_B$/f.u. at 2 nm of film thickness. Figure 2 shows magnetic properties of epitaxial $ErFeO_3$ thin films at 300 K, with varying film thickness from 1.2 to a few tens of nm. We assume that only $Fe^{3+}$ ions are responsible for the observed magnetization at 300 K, as $Er^{3+}$-$Er^{3+}$ exchange interaction is likely to be activated at much lower temperatures (4.5 K in bulk). Figure 2a-d show $M(H)$ hysteresis curves from magnetic property measurement system (MPMS) results of $ErFeO_3$ thin films with film thicknesses of 1.2, 2, 3, and 4 nm, with an external magnetic field swept between −5 and 5 T, along three orthogonal directions. Magnetic hysteretic behavior is extremely weak, with the remnant magnetization on the order of ~0.1 $\mu_B$/f.u., as shown in the insets of Figure 2a-d, consistent with the weak canted ferromagnetic behavior. For all the samples, magnetization steeply increases and saturates to a maximum value of $M_s$ at around 1 T of magnetic field. In addition, compared to the negligible 90° in-plane magnetic anisotropy between the red solid and dashed lines, the out-of-plane magnetization (orange line) is significantly larger than the in-plane magnetizations. This magnetic anisotropy originates from structural parameters such as shape (dimension), epitaxial strain, octahedral rotation/tilt, and domain formation. Surprisingly, $M_s$, which is a magnetic response to the external magnetic field, increases as the thickness decreases, reaching a maximum value of 1.5 and 3.7 $\mu_B$/f.u. above 1 T along the in-plane and out-of-plane directions, respectively, for a 2-nm-thick film. This value is approximately two orders of magnitude larger than the bulk value of ~0.04 $\mu_B$/f.u. at low temperatures (in bulk, $ErFeO_3$ is antiferromagnetic with effectively zero $M_s$ at 300 K) [29]. The increase of $M_s$ with decreasing film thickness is against conventional wisdom, where the magnetic ordering becomes more difficult to stabilize as the spatial dimension decreases. For example, a ferrite ($Fe_3O_4$) thin film on MgO substrate shows decreasing paramagnetic response and $M_s$ as the film thickness decreases [30]. Ferromagnetic properties, including the transition temperature and spontaneous magnetization, also decrease significantly with decreasing film thickness in the $SrRuO_3$//$SrTiO_3$, $SrRuO_3$//Nb:$SrTiO_3$, and $La_{0.7}Sr_{0.3}MnO_3$//$SrTiO_3$ systems and vanish at the ultrathin limit [31-33]. This result implies that the dimensional effects such as quantum confinement and/or Mermin-Wagner fluctuation, which determine low-dimensional magnetism in conventional systems, are effectively suppressed or deliver different magnetic responses at least down to a 2-nm-thick $ErFeO_3$ film. This giant, thickness-dependent magnetic response is consistently obtained during PNR analyses, despite some quantitative differences between MPMS and PNR results. One possible origin for the quantitative difference between MPMs and PNR results is the sample size effect.

We note that below the $ErFeO_3$ film thickness of 2 nm, $M_s$ is suppressed. This may result from the activation of Mermin-Wagner fluctuations at the 2D limit or structural instability for the few-unit-cell-thick film, leading to unstable DMI. Depth profiles of in-plane magnetization show that the magnetic response is maximized for the 2-nm-thick $ErFeO_3$ film, consistent with the MPMS measurements. We extracted layer-by-layer magnetization from the difference between spin-up and spin-down neutron reflectivity in polarized neutron reflectivity (PNR) measurements [34-40]. Upper panels of Figure 2e-g show PNR results of 2, 4, and 7 nm-thick $ErFeO_3$ films, respectively, at 300 K with a 4.5 T magnetic field. The red and blue symbols are experimental reflectivity of spin up ($R^+$) and down neutrons ($R^-$), respectively. The solid lines of corresponding colors represent reflectivity modeling results using thickness, roughness, and in-plane magnetic moment as fitting parameters. The insets in Figures 2e-g show magnified views of the 0.6-0.8 $nm^{-1}$range serving to more clearly illustrate the differences between $R^+$ and $R^-$. The obtained results indicate the presence of a small yet finite difference, which increases as the film thickness decreases. The lower panels of Figures 2e-g present the experimental and simulated spin asymmetry, defined as the normalized difference between $R^+$ and $R^-$. By fitting the calculated spin asymmetry to the experimental data, depth profiles of the nuclear and magnetic scattering length density of the thin films were obtained, as shown in Figure 2h-j. The results of fitting the scattering length density profiles shows that 2, 4, and 7 nm-thick films exhibit the same nuclear scattering density of approximately $6 \times 10^{-6}$ Å$^{-2}$, for the thin film region, but varying magnetic scattering length density of $0.33 \times 10^{-6}$, $0.18 \times 10^{-6}$, and $0.15 \times 10^{-6}$ Å$^{-2}$, respectively. This magnetic scattering length density corresponds to average magnetic moments oriented within the plane of the film, with magnitudes of 0.72, 0.36, and 0.32 $\mu_B$/f.u. for the 2, 4, and 7 nm-thick films, respectively. Figure 2k summarizes the magnetic response of the $ErFeO_3$ epitaxial thin film as a function the film thickness. The $M_s$ value of $ErFeO_3$ thin film increases as the film thickness decreases, down to 2 nm, where a distinct peak corresponding to the maximum $M_s$ value is observed. Due to the small neutron flux, samples for PNR have been grown on 10 mm × 10 mm size substrates, which are 4 times larger than those for MPMS. The different sample sizes may lead to different shape anisotropy, domain structure, and/or edge effects during the PLE growth and magnetization measurements.

The magnetic responses of epitaxial $ErFeO_3$ thin films at low temperature also exhibit the same film thickness-dependent trends, but with significantly higher magnetization values, partly attributed to the long-range order of $Er^{3+}$ ions. Figure S3 shows temperature-dependent magnetization ($M(T)$) results with a 0.1 T magnetic field, suggesting that 2-nm-thick film exhibits the strongest magnetic response—evidenced by higher magnetization in the temperature range of ~25-300 K, particularly in the direction perpendicular to the film plane (out-of-plane)—against the background of the diamagnetic contribution from the $SrTiO_3$ substrate. Figures S4a-d show the MPMS results at 2 K. The $M_s$ increases as the film thickness decreases, reaching maximum values of 4.9 and 8.5 $\mu_B$/f.u., along the in-plane and out-of-plane directions, respectively, for the 2 nm-thick film. While films with thicknesses of 3, 4, and 20 nm exhibit weakly pronounced hysteretic behavior with negligibly small remnant magnetization, the 2 nm-thick film is characterized by a remnant magnetization of 0.2 and 1.1 $\mu_B$/f.u, along in-plane and out-of-plane directions, respectively, which significantly exceeds the values obtained at 300 K. Furthermore, films with thicknesses of 3, 4, and 20 nm reach $M_s$ at a magnetic field of ~3 T of, which is also larger than the saturation value observed at ~1 T and a temperature of 300 K. The upper panels of the Figure S4e-g show the PNR results of 2, 4, and 7 nm-thick films, respectively, obtained at a temperature of 1.5 K and a magnetic field of 4.5 T. A clear difference between $R^+$ and $R^-$ is shown which increase as the film thickness decreases. The lower panels of Figure S4e-g present experimental and calculated spin asymmetry, demonstrating good agreement for all measured samples. Depth profiles of the nuclear and magnetic scattering length densities of the thin films are shown in Figure S4h-j. The nuclear scattering length densities of the thin films were obtained to be $6 \times 10^{-6}$ Å$^{-2}$, consistent with the value obtained at 300 K. The magnetic scattering length densities of the 2, 4, and 7 nm-thick films were obtained to be $3.0 \times 10^{-6}$, $2.3 \times 10^{-6}$, and $1.8 \times 10^{-6}$ Å$^{-2}$, which corresponds to 6.5, 5.0, and 4.0 $\mu_B$/f.u. of average in-plane magnetic moments, respectively. Both magnetization measurements and PNR analyses consistently demonstrate that the magnetic response intensifies as the film thickness decreases, even at low temperatures. Figure S5 shows the PNR results at intermediate temperatures. Fitting of the PNR spectra obtained at 40, 100, and 150 K yields an in-plane magnetic moment of 2.0 $\mu_B$/f.u. This value exceeds the 0.72 $\mu_B$/f.u. observed at 300 K but falls short of the 6.5 $\mu_B$/f.u. observed at 1.5 K. This suggests that long-range ordering of $Er^{3+}$ ions is responsible for the substantial magnetic enhancement at low temperature. Figure S6 shows Er *M*-edge XMCD data at 12.85 K, which demonstrates the finite spin ordering of the $Er^{3+}$ moments. This serves as further confirmation of the simultaneous spin ordering of both $Er^{3+}$ and $Fe^{3+}$ ions, which governs the low-temperature magnetic behavior of the system.

To understand the giant magnetic enhancement at the ultrathin limit, we focused on the octahedral tilt pattern near the interface of the $ErFeO_3$ film and $SrTiO_3$ substrate using STEM. Figure 3a presents a STEM image of a ~13-nm-thick $ErFeO_3$ film projected along the $[110]_{pc}$ zone axis. The film exhibits alternating 90° twin domains of the orthorhombic $ErFeO_3$ structure, oriented along the [010] (red) and [100] (green) directions. The insets in Figure 3a show the atomic models of both orientations superimposed on magnified, inverted annular bright-field (ABF) images, which clearly resolve the oxygen columns. These atomically sharp 90° domain boundaries arise from nearly energetically degenerate strain states along the two orthogonal axes of the $ErFeO_3$ grown on cubic $SrTiO_3$ substrate. The characteristic octahedral distortion can be quantified by the tilt angle, $\theta$, along the [010] zone axis, representing the angular deviation of the Fe–O–Fe bond from 180°. To evaluate the spatial distribution of these rotation angles along the growth direction, an inverted ABF image was acquired across the interface within a uniform [010] region (Figure 3b). The corresponding octahedral rotation map (Figure 3c), determined via precise atomic-column fitting of the oxygen positions, confirms that the periodic alternation of the Fe–O–Fe chains is continuous and extends roughly 2–3 u.c. into the $SrTiO_3$ substrate. However, a pronounced suppression of the rotation angle is observed directly at the $ErFeO_3$/$SrTiO_3$ interface. This structural evolution is quantitatively illustrated by the depth profile of $\theta$ in Figure 3d, where $\theta$ approaches its bulk value of ~20° within the interior of the film [41]. Approaching the interface—specifically within the first two unit cells above the epitaxial plane—the value of $\theta$ decreases significantly to 13° and 6°. Concurrently, the domain structure becomes degenerate just above the interface. This observation is highlighted in Figure 3a, where the interface region (2 u.c. layer) is shown in yellow dashed boxes in Figure 3a. To confirm the robustness of this behavior, an ultrathin ~2-nm-thick film was investigated. As shown in Figure S7, a remarkably similar interfacial octahedral rotation profile is preserved. The reproducibility of this domain arrangement across multiple regions of the ~2-, and ~13 nm-thick film is further demonstrated in Figures S8 and S9.

The drastic strain relaxation behavior at the interface of $ErFeO_3$ ultrathin film is further confirmed for a larger length scale. Figure 3e-h show XRD measurements on several half-order Bragg diffraction peaks that correspond to unit-cell doubling from each octahedral tilt pattern [42,43]. In Figure 3e,f, the two representative reciprocal space regions of a 2 nm-thick $ErFeO_3$ film for half-order analyses, (3/2, 1/2, L), (1/2, 3/2, L), respectively, exhibit only one half-order peak (1/2, 3/2, 3/2) corresponding to $b^-$ or $c^-$ tilt. On the other hand, in Figure 3g,h, the same reciprocal space regions of 30 nm-thick film show several half-order peaks including (3/2, 1/2, 2/2), (3/2, 1/2, 3/2), (3/2, 1/2, 5/2), (1/2, 3/2, 2/2), and (1/2, 3/2, 3/2) corresponding to the combined contribution of $a^-$, $b^-$, $c^-$, and $c^+$ tilt. These results confirm that 2 nm-thick $ErFeO_3$ film exhibit $a^0a^0c^-$ octahedral distortion with suppressed tilt, whereas 30 nm-thick $ErFeO_3$ film exhibit $a^+b^-b^-$ octahedral distortion with predominant octahedral tilting. The thickness-dependent transition of octahedral tilt from XRD can be further interpreted as macroscopic confirmation of the suppressed tilt pattern and consequent suppression of the orthorhombic domain formation for the ultrathin film observed in STEM images. More detailed half-order peak analyses of XRD results are shown in Figure S10, which support the same conclusion. Suppression of structural domain formation due to the straightened octahedral tilt near the interface may contribute to the enhanced magnetic response of $ErFeO_3$ at the ultrathin limit. As shown in Figure 3 and Figures S7-S9, the [100] and [010] structural domains alternate along the in-plane direction in $ErFeO_3$ films.

Earlier reports suggest that the domain walls may enhance net magnetic moment of an antiferromagnet, producing an uncompensated antiferromagnetic moment [44,45]. However, considering the structural and chemical quality, $ErFeO_3$ epitaxial thin films exhibit minimized extrinsic defects with a dominant $Fe^{3+}$ valence state. The results of magnetic measurements also rule out the role of defects at the domain walls contributing to the magnetization, because $M_s$ increases as the film thickness and the volume fraction of the domain walls decreases. In contrast, the domain walls may function as magnetic pinning centers which hinder magnetic saturation [46]. Since the lattice structure abruptly shifts at the domain walls, the exchange interaction for long-range spin order decreases quickly, and magnetic pinning centers may be introduced. The magnetic pinning effect suppresses the magnetic response of thicker $ErFeO_3$ films with substantial domain structures. Because the structural domains merge at the interface, the magnetic response of 2 nm-thick film suffers less from the magnetic pinning effect. Figure 3i shows a schematic image demonstrating that the magnetic pinning by domain walls is predominantly active in the bulk thin-film region away from the interface. The observed quantitative difference in magnetization values between the in-plane and out-of-plane directions can be explained by the in-plane domain wall structure responsible for larger magnetization suppression along the in-plane direction. In addition, the intrinsic magnetic phase $\Gamma_4(G_x, A_y, F_z)$ of $ErFeO_3$ at 300 K exhibits a weak ferromagnetic moment along the out-of-plane direction in the first place. Hence, the defect-free domain wall structure may explain the large enhancement of the magnetization in the ultrathin limit of $ErFeO_3$. To support this magnetic domain scenario, Figure S11a,b further present magnetic force microscopy (MFM) results that directly visualize the magnetic domains under an out-of-plane magnetic field sweeping from +4 to −4 T at the surface of the 2 nm-thick film at 4.2 and 20 K respectively. The MFM image confirms that at both temperatures the magnetic domains are observed at the surface, and the contrast of frequency shift, indicating a magnetic gradient, between the magnetic domains is small at low magnetic fields and grows with increasing magnetic field. The comparison between MFM images at 4.2 and 20 K shows that the overall range of frequency shift at 4.2 K is between 0.25 and −0.25 Hz, which is larger than between 0.2 and −0.2 Hz at 20 K. It suggests that the magnetic gradient between magnetic domain can be more distinct at 4.2 K than at 20 K. as expected.

The distinct magnetic contribution of the characteristic domain wall structure of $ErFeO_3$ does not contradict the depth profile of the spin ordering obtained from the PNR analyses. Instead of a uniform magnetic distribution within a layer, we may consider different magnetic responses from the interface and bulk thin-film region. Figure S12a-c shows the simulated spin asymmetry for the uniform magnetic distribution (as shown in Figure 2e-g), compared to a bilayer model with a larger magnetic moment near the interface, for the 2, 4, and 7 nm-thick films, respectively. Figure S12d-f show the scattering length density results of the newly introduced bilayer model for the corresponding samples. The 2, 4, and 7 nm-thick films may have ~0.8 nm-thick layer at the interface with 1.5, 0.6, and 0.5 $\mu_B$/f.u. of magnetization, 1.2, 3.2, and 6.2 nm layers above the interfacial layers with 0.1, 0.3, and 0.3 $\mu_B$/f.u., respectively. Since both simulations match well with the experimental spin asymmetry, the layer model with a larger magnetic moment residing near the interface serves as a complementary magnetic structure. This makes the domain effect consideration more relevant.

Orbital reconstruction at the surface of orthoferrites may substantially reduce the antiferromagnetic exchange interaction, to further assist the spin ordering at the ultrathin limit. In addition to the structural domain picture, a modification of the local exchange interaction near the film surface offers a complementary, microscopic understanding of why the field-induced magnetization grows as the $ErFeO_3$ film becomes atomically thin. For this, we performed first-principles density functional theory (DFT) calculations on $FeO_2$-surface-terminated $ErFeO_3$/$SrTiO_3$ films (see Methods; an extended polar-discontinuity scenario and calculation results on the complementary superlattice geometry are presented in Notes S1 and S2, respectively). A substantial reconstruction of the Fe $e_g$ orbital polarization at the surface is shown in Figure 4. The layer-resolved Fe $d$-projected density of states reveals that bulk-like Fe ions carry a predominantly $d_{z^2}$-character at the top of the valence band (Figure S14a), whereas the topmost (surface) Fe ion, which lacks an apical oxygen on the vacuum side, instead exhibits a clear $d_{x^2-y^2}$-character (Figure 4a). Consequently, the vertical nearest-neighbor superexchange between the surface Fe and the Fe directly beneath it proceeds between two orthogonal $e_g$ orbitals ($d_{x^2-y^2}$ and $d_{z^2}$) rather than between the parallel $d_{z^2} \leftrightarrow d_{z^2}$ orbitals of the bulk. By the Goodenough–Kanamori rule, this orthogonal-orbital channel is only weakly antiferromagnetic — or even ferromagnetic — in contrast to the strongly antiferromagnetic $d_{z^2} \leftrightarrow d_{z^2}$ channel of the bulk, so that the surface hosts a local tendency toward ferromagnetic alignment of the Fe spins. Because the reconstruction is confined to the topmost Fe layer, its relative weight increases as the film thins toward an ultrathin limit and the surface Fe constitutes a larger fraction of all Fe sites, qualitatively consistent with the experimentally observed enhancement of the field-induced magnetization toward the 2-nm-thick film. Figure 4f summarizes $\Delta E = E_{FM} - E_{G\text{-}AFM}$ in eV/Fe as a function of the unit cell thickness. As the value decreases, the relative stability of ferromagnetic phase can be inferred. The experimental octahedral distortion pattern away from the interface can be described as an S2 structure with both rotation/tilt patterns, which shows a minimum $\Delta E$ for 4 u.c.-thick $ErFeO_3$ film, qualitatively consistent with the maximum $M_s$ observed experimentally. However, the suppression of magnetization for the films thinner than 2 nm, which is not covered by the orbital reconstruction, requires an alternative theoretical approach. The polar discontinuity scenario in Note S1 and Figure S13 suggests that polar band bending reduces with decreasing thickness, which induces recovery of the antiferromagnetic stability below the thickness of 4 u.c. Note that the reconstruction originates purely from the surface bonding environment and is accompanied by no appreciable change in the Fe valence, consistent with the exclusively $Fe^{3+}$ assignment of the experimental Fe *L*-edge XAS. While this picture needs to be taken as a qualitative argument, we suggest that the surface engineering of orbital polarization as observed here might be another pathway to harness magnetism for the realization of ultrathin ferromagnetism in the interfacial area (see Methods, Notes S1 and S2 for more details).

## 3. Conclusion

Our study demonstrates that giant enhancement of magnetic response in $ErFeO_3$ thin film is realized in the ultrathin limit with maximum $M_s$ value of 3.7 $\mu_B$/f.u. at 300 K for 2 nm-thick film. This value is substantially higher than that of bulk $ErFeO_3$, which is an antiferromagnet at room temperature. Structural and magnetic analyses suggest that the giant enhancement of magnetic response in ultrathin film is induced by suppression of domain formation stemming from epitaxial strain, which excludes the magnetic pinning effect at the orthorhombic domain wall. The insight from the theoretical approach further supports magnetic enhancement in ultrathin films with the orbital reconstruction model, which leads to substantial differences in exchange interaction between the film surface and subsurface. This result gives rise to less AFM even FM-like coupling by Goodenough-Kanamori rule. Considering that the two key mechanisms are promoted by epitaxial strain and surface structure, which are only accessible by thin film geometry, we propose the design of synthetic thin-film structure with atomic-scale precision as an effective platform for realizing low-dimensional magnetism for spintronic and memory application at 300 K, which is practical condition for functional magnetic application.

## 4. Experimental Methods

**4.1 Sample growth:** $ErFeO_3$ thin film is grown on $SrTiO_3$ (001) substrates (Shinkosha) by pulsed laser epitaxy using KrF laser (248 nm, IPEX-868, Lightmachinary). The $TiO_2$ surface termination of the $SrTiO_3$ substrates were prepared by treatment with buffered HF and annealing at 1000 °C for 6 h. We fabricated films at 750 °C with oxygen partial pressure of 100 mTorr, laser fluence of 1.5 J $cm^{-2}$, and repetition rate of 5 Hz.

**4.2 X-ray diffraction (XRD):** Synchrotron XRD at 8D beamline of the Pohang light source (PLS) is performed to confirm the thickness and quality of $ErFeO_3$ ultrathin film. X-ray reflectivity and $\theta$-$2\theta$ scan confirms the thickness and quality of the films. Synchrotron XRD at PLS 5D beamline is performed to confirm octahedral tilt via XRD half-order analyses. Homelab high-resolution XRD (X'Pert PRO, Malvern Panalytical) with Cu Ka radiation is also performed to reproduce thicker samples and reciprocal space mapping.

**4.3 Scanning transmission electron microscopy (STEM):** Cross-sectional TEM lamellae were prepared using a focused ion beam (FIB, Helios 5, Thermo Fisher Scientific, USA). The FIB-prepared specimens were mounted on Cu grid for STEM characterization. In the final thinning step, a low energy $Ga^+$ ion beam (2 kV) was used to minimize ion-beam–induced damage. STEM imaging was carried out in a probe-corrected field-emission transmission electron microscope (Spectra Ultra 300, Thermo Fisher Scientific, USA) in scanning mode operating at 300 kV. Atomic-scale characterization was recorded in annular bright-field (ABF), high-angle annular dark-field (HAADF) imaging modes. The imaging modes were set with detector angle ranges of 7.5 to 17 mrad for ABF, 52 to 200 mrad for HAADF. The convergence semi-angle for forming the focused probe was maintained at 21.5 mrad. Energy-dispersive X-ray spectroscopy (EDS) was performed with a detector of about 4 srad solid angle. Probe current was set about 70 pA. To measure octahedral rotation/tilt, oxygen positions were determined from inverted ABF images using center-of-mass and two-dimensional Gaussian fitting within the open-source software Atomap [47].

**4.4 X-ray absorption spectroscopy (XAS):** XAS, X-ray magnetic circular, linear dichroism, (XMCD, XMLD) of 2 nm-thick $ErFeO_3$ Fe $L_{2,3}$-edges are measured at the 6A beamline of the PLS in total electron yield mode. Measurements are conducted with a 45° incident angle, 0 and 7 T of magnetic field along the X-ray direction, at 3 and 300 K. XMCD and XMLD are obtained by calculating the difference between left and right circular polarization and horizontal and vertical polarization, respectively. Low-temperature Er $M_{4,5}$ edges XMCD measurements are performed at BL13-3 of Stanford Synchrotron Radiation Lightsource (SSRL). The sample was mounted on top of a permanent magnet and cooled down to the base temperature (12.85 K).

**4.5 Magnetic property measurement system (MPMS):** Field dependence of magnetization $M(H)$ data was measured with horizontal and perpendicular 5 T field at 300 and 2 K MPMS (Quantum Design) to confirm thickness dependence of $M_s$. Magnetization is normalized by the total number of Fe ions, which is calculated from the total volume of the film 5 mm × 5 mm × (film thickness) divided by perovskite unit cell volume 3.796 Å × 3.905 Å × 3.905 Å.

**4.6 Polarized neutron reflectivity (PNR):** PNR experiments were performed on the Magnetism Reflectometer MAGREF at the Spallation Neutron Source at Oak Ridge National Laboratory [38,48-50], using neutrons with wavelengths λ in the range of 0.2–0.8 nm and a high degree of polarization of 98.5–99%. The measurements were performed in a closed cycle refrigerator equipped with a 5 T cryomagnet (Cryomagnetics). Using the time-of-flight method, a beam of collimated polychromatic polarized neutrons with a wavelength band of δλ impinged onto a film at a grazing angle θ, interacting with atomic nuclei and the spins of unpaired electrons. The reflected intensity $R^+$ and $R^-$ was measured as a function of momentum transfer, $q = 4\pi \sin(\theta)/\lambda$, with the neutron spin parallel (+) or antiparallel (−) to the applied field, respectively. To separate nuclear and magnetic scattering, the spin asymmetry ratio $SA = (R^+ - R^-)/(R^+ + R^-)$ was calculated, where SA = 0 corresponds to a zero magnetic moment in the system. Being electrically neutral, spin-polarized neutrons penetrate the entire multilayer structures and probe magnetic and structural composition of the film and buried interfaces all the way down to the substrate.

**4.7 Magnetic force microscopy (MFM):** MFM measurements were performed using a custom-built magnetic force microscope equipped with a 2–2–9 T vector magnet [51]. All measurements were carried out using a commercial MFM tip (PPP-MFMR, NANOSENSORS). The MFM images were acquired at 4.2 and 20 K while sweeping the out-of-plane magnetic field from +4 to −4 T, under the following imaging conditions: a scan area of 2.5 × 2.5 $\mu m^2$, a resolution of 128 × 128 pixels, and a tip–sample distance of 100 nm.

**4.8 First-principles calculations:** Spin-polarized density functional theory calculations were performed for $FeO_2$-terminated $[ErFeO_3]_n/[SrTiO_3]_2$ films separated by a vacuum region and $[ErFeO_3]_n/[SrTiO_3]_2$ periodic superlattices, with $n$ = 1 to 6 . The in-plane lattice constant was fixed to the cubic $SrTiO_3$ value ($a_{STO}$ = 3.905 Å) to enforce the experimental epitaxial strain, while the out-of-plane lattice parameter and all internal atomic coordinates were relaxed. For the film geometry, an SrO bottom termination and an $FeO_2$ top termination were used unless otherwise specified. Four collinear magnetic configurations (FM, A-AFM, C-AFM, G-AFM) and three $FeO_6$ rotation/tilt patterns (Str. 1: $a^0a^0c^-$; Str. 2: $a^+b^-b^-$; Str. 3: $a^0a^0c^+$) were sampled at every thickness to locate the magnetic and structural ground state. In the manuscript layer-resolved projected density of states from the $n = 2$ film in the lowest-energy state (Str. 3 + G-AFM). We employed the Vienna ab initio Simulation Package (VASP) code [52-56]. The generalized gradient approximation in the Perdew-Burke-Ernzerhof parametrization for solids (PBEsol) is adopted [57]. 500 eV of plane-wave energy cutoff and 6 × 6 × 1 of the Monkhorst-Pack k-grid sampling were used. In addition, the Fe 3*d*-orbitals were described using the on-site repulsive correction via the rotationally invariant DFT + $U$ approach [58], with the effective Hubbard $U$-parameter $U_{eff} = 4$ eV was employed. Structural relaxations of internal coordinates were performed via employing 1 meV / Å of force criteria with the in-plane lattice parameter fixed (in the superlattice cases, the out-of-plane lattice parameters were also optimized). In the film geometry, 20 Å of vacuum was placed in all unit cells.

**Acknowledgements**

This work was supported by the National Research Foundation of Korea (NRF) grant funded by the Ministry of Education, Science and Technology of Korea government (MSIT) (Nos., RS-2026-25488300, RS-2023-00220471, RS-2026-25486986, RS-2026-25517484, RS-2022-NR068223, RS-2024-00410027, and RS-2026-25472018). The TEM work at KENTECH was supported by the KENTECH Center for Shared Research Facilities. A portion of this research used resources at the Spallation Neutron Source, a DOE Office of Science User Facility operated by the Oak Ridge National Laboratory. The beam time was allocated to BL-4A, Magnetism Reflectometer, on proposal numbers IPTS-28438.1 and IPTS34480.1. Experiments at PLS-II were supported in part by MSIT and POSTECH. XRD experiments were conducted at 5D and 8D beamlines of PLS, and XAS experiments were conducted at the 6A beamline of PLS. Low-temperature XMCD measurements were carried out at the Stanford Synchrotron Radiation Lightsource (SSRL), SLAC National Accelerator Laboratory, which is supported by the US Department of Energy, Office of Science, Office of Basic Energy Sciences (contract No. DE-AC02- 76SF00515). The United States Government retains, and the publisher, by accepting the article for publication, acknowledges that the United States Government retains, a nonexclusive, paid-up, irrevocable, worldwide license to publish or reproduce the published form of this manuscript, or allow others to do so, for United States Government purposes. The Department of Energy will provide public access to these results of federally sponsored research in accordance with the DOE Public Access Plan (http://energy.gov/downloads/doe-public-access-plan)

**Funding**

This work was supported by the National Research Foundation of Korea (NRF) grant funded by the Ministry of Education, Science and Technology of Korea government (MSIT) (Nos. RS-2026-25488300, RS-2023-00220471, RS-2026-25486986, RS-2026-25517484, RS-2022-NR068223, RS-2024-00410027, and RS-2026-25472018).

**Conflicts of Interest**

The authors declare no conflicts of interest.

**Data Availability Statement**

The data that support the findings of this study are available in the supporting information of this article.

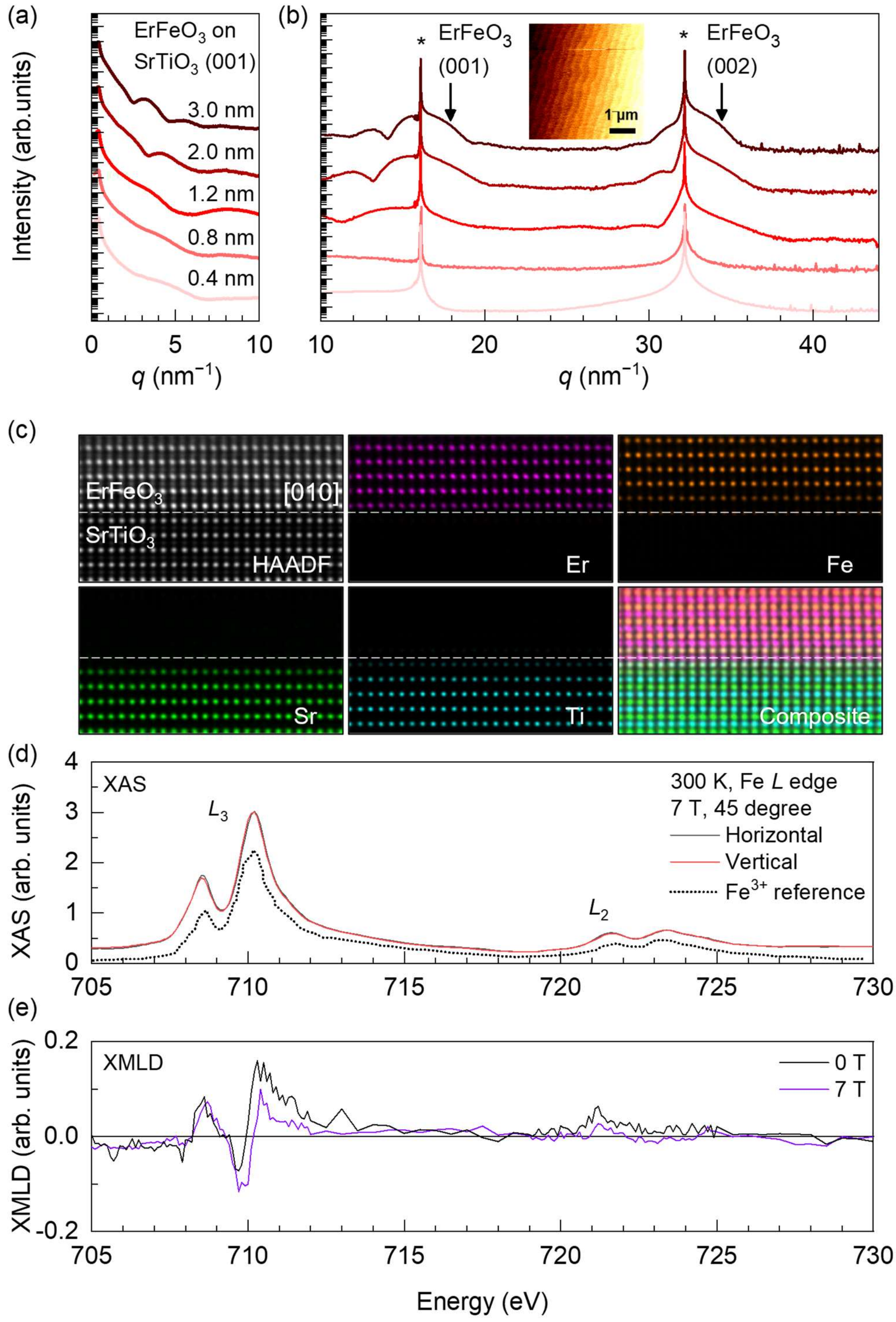

**Figure 1.** (a) X-ray reflectivity and (b) X-ray diffraction $\theta$-$2\theta$ scan results of $ErFeO_3$ ultrathin films on $SrTiO_3$ substrates. The film and substrate peaks are denoted as arrows and stars, respectively. The inset shows the atomic force microscopy image of the film. (c) Atomic resolution image of high-angular annular dark-field scanning transmission electron microscopy, and energy dispersive spectroscopy of Er, Fe, Sr, Ti, and the composite. Fe $L_{2,3}$-edges X-ray absorption spectroscopy of 2-nm-thick $ErFeO_3$ film for (d) horizontal and vertical polarizations and $Fe^{3+}$ reference. The measurement was conducted with 45° incident angle, 7 T of magnetic field along X-ray direction, at 300 K, (e) X-ray magnetic linear dichroism with 0 and 7 T of magnetic field.

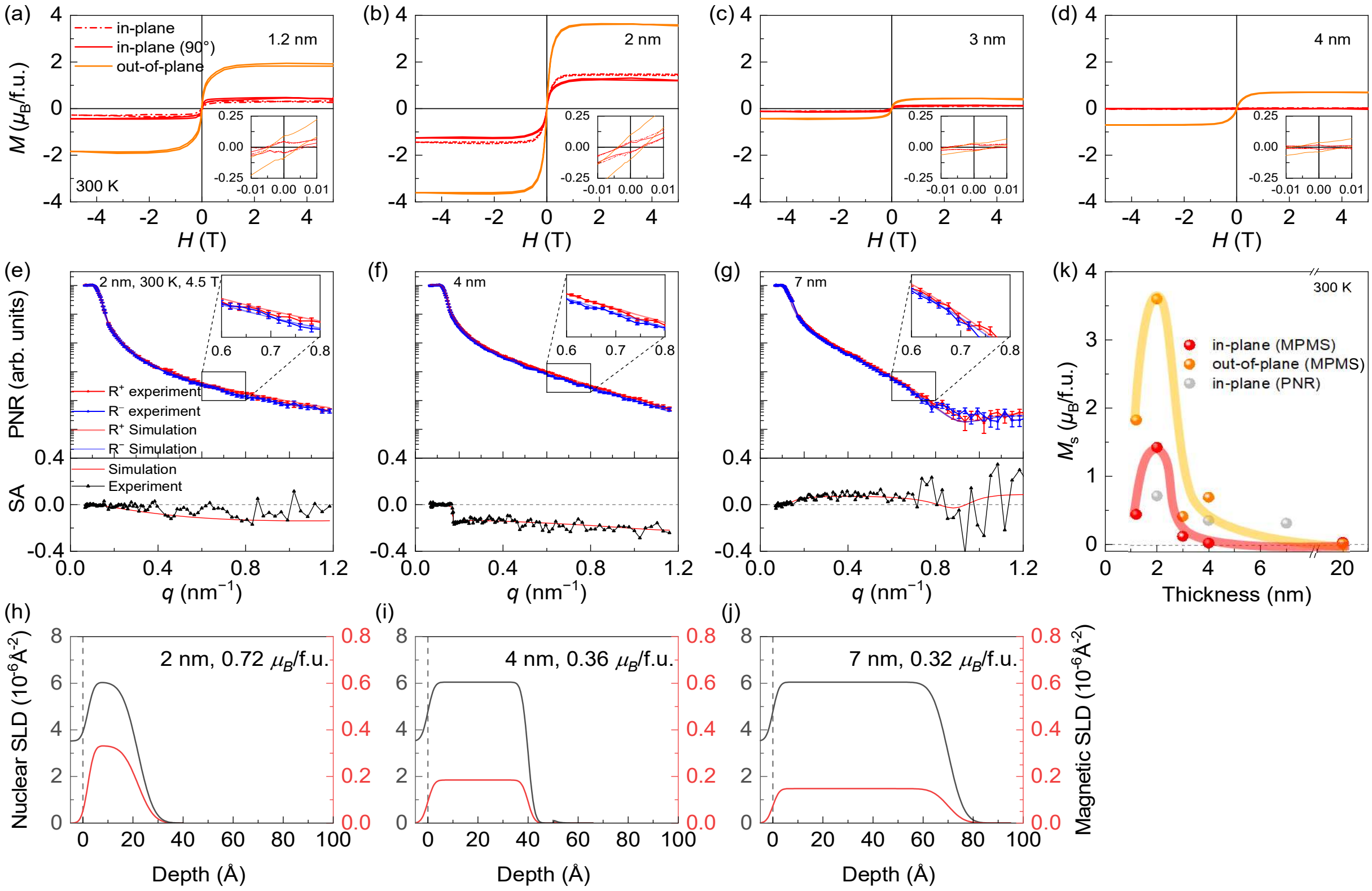

**Figure 2.** (a-d) Magnetic property measurement system data of 1.2, 2, 3, 4 nm $ErFeO_3$ films, insets focus on the regions near $H = 0$, (e-g) Upper panels: polarized neutron reflectivity of 2, 4, 7 nm $ErFeO_3$ film at 300 K with 4.5 T, Lower panels: experimental and simulation spin asymmetry. (h-j) Nuclear, magnetic scattering length density of (e-g). Vertical gray dash line indicates $ErFeO_3/SrTiO_3$ interface (k) Set of saturation magnetization values with respect to film thickness, from magnetic property measurement system, and polarized neutron reflectivity.

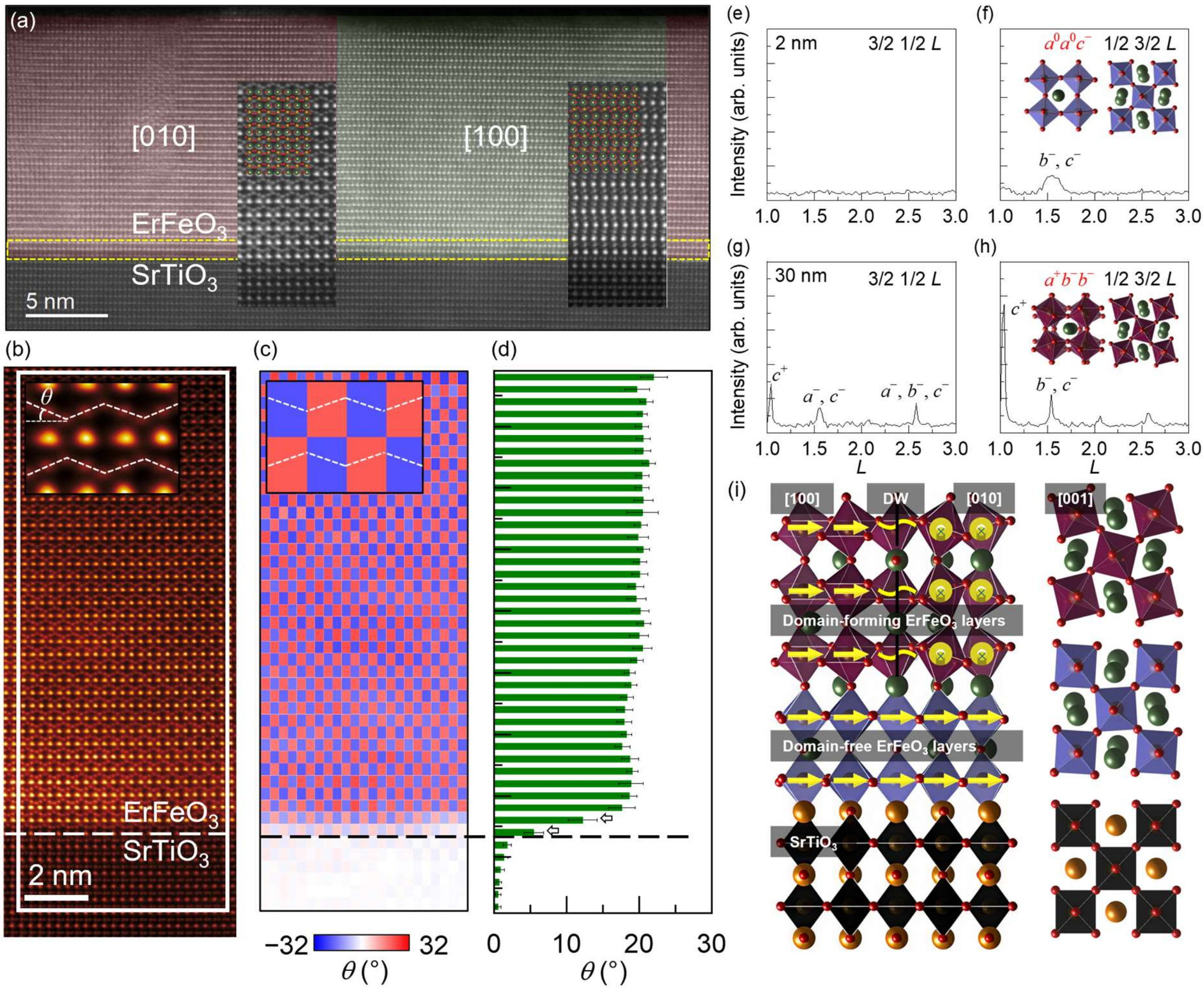

**Figure 3.** (a) Scanning transmission electron microscopy plot of $ErFeO_3$ thin film, two different domains are shown. (b) Scanning transmission electron microscopy plot of only [010] domain and the definition of tilt-angle $\theta$ of the domain. (c) Color mapping of $\theta$ and (d) $\theta$ values of each layer. (e,f) X-ray diffraction half-order measurement of 2 nm $ErFeO_3$ (g,h) X-ray diffraction half-order measurement of 30 nm $ErFeO_3$, (i) schematic diagram of domain structures of $ErFeO_3$ thin film and magnetic pinning effect.

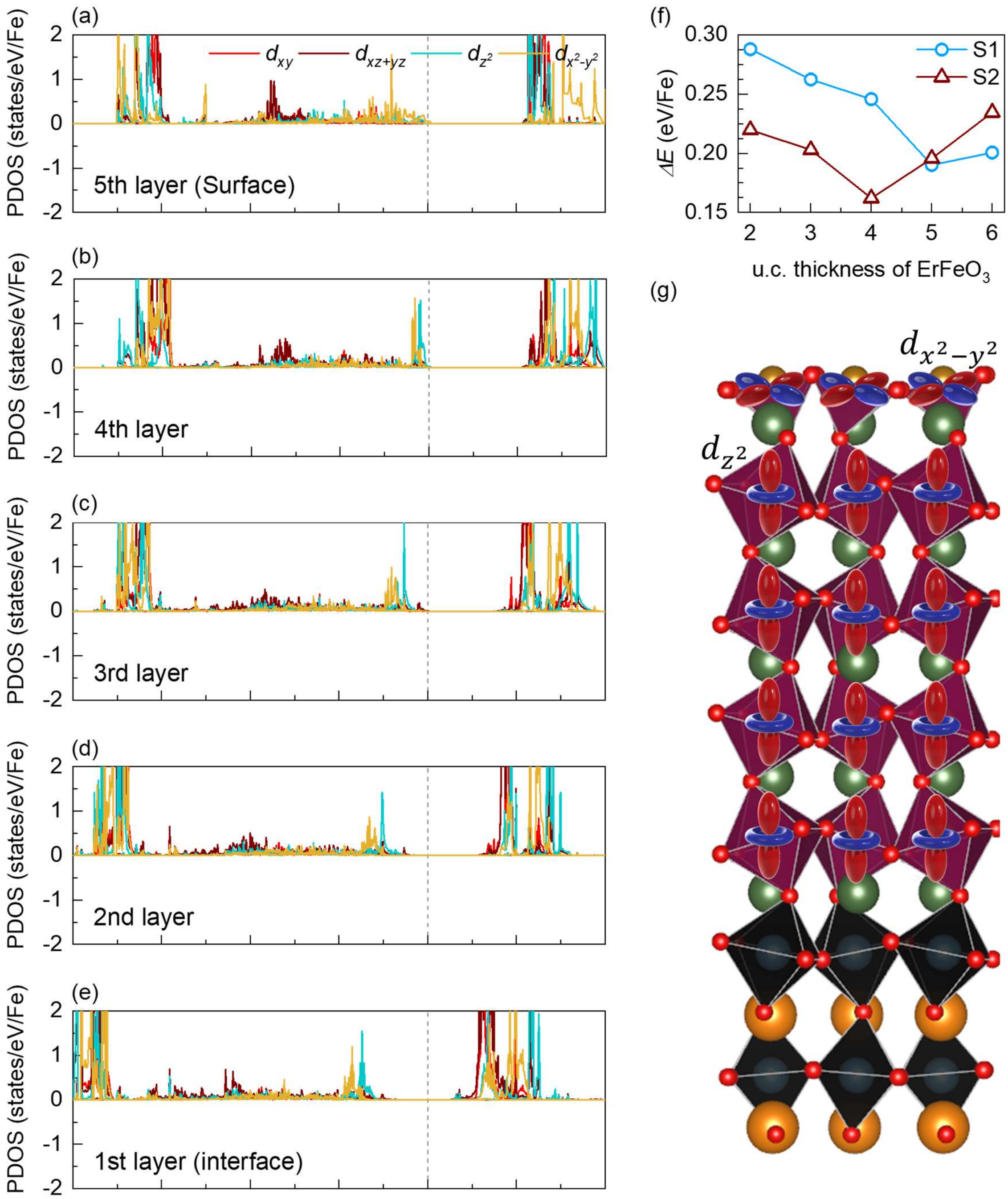

**Figure 4.** Surface $e_g$-orbital-polarization reconstruction in the surface-terminated $ErFeO_3$ film. (a) Layer-resolved Fe *d*-orbital-projected density of states for (a) the topmost surface Fe, showing dominant $\boldsymbol{d_{x^2-y^2}}$ character driven by the missing apical oxygen on the vacuum side, and (b-e) Fe of subsurface layers which shows bulk-like property. (f) Unit cell thickness-dependent magnetic stability of the $ErFeO_3/SrTiO_3$ film. Calculated $\Delta E = E_{FM} - E_{G\text{-}AFM}$ (in eV/Fe) versus $ErFeO_3$ unit cell thickness, for the microscopic structures that has (S1) only octahedral rotation ($a^0a^0c^-$) and (S2) both octahedral rotation and tilt ($a^+b^-b^-$) distortions. (g) Schematic diagram highlights the change of the orbital polarization of the surface Fe.

Supporting Information

# Giant Magnetization Enhancement at Ultrathin Limit of an Orthoferrite

*Jin Young Oh, Yaolong Xing, Seung Gyo Jeong, Sehwan Song, Woo-suk Noh, Valeria Lauter, Seongjae Choi, Seyoung Kwon, Jaeha Choi, Si Won Kim, Hong Hyun Jeon, Heemin Lee, Cheng-Tai Kuo, Jun-Sik Lee, Chanyoung Lee, Yeonkyu Lee, Jeehoon Kim, Heung-Sik Kim, Sungkyun Park*, Sang Ho Oh*, Woo Seok Choi**

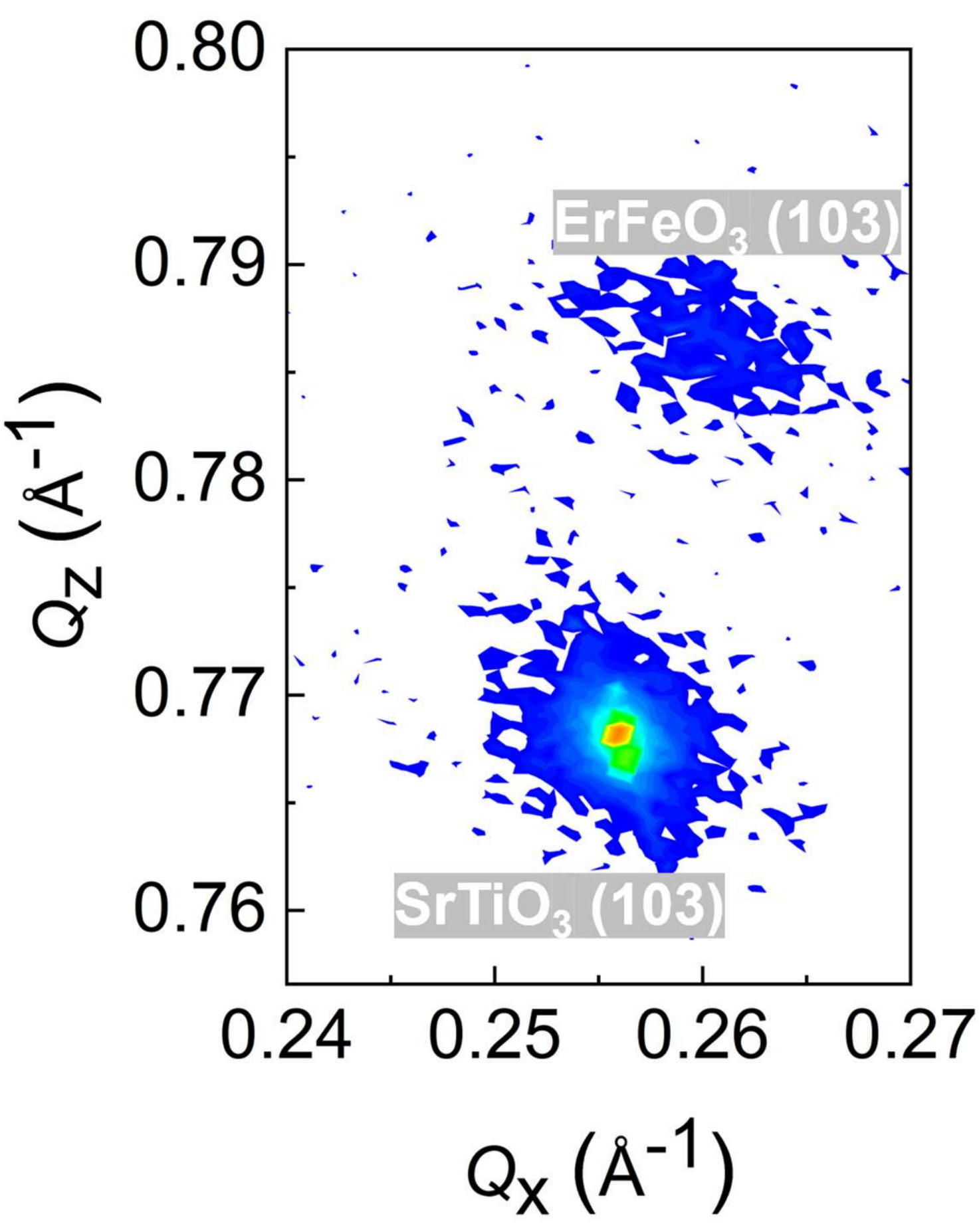

**Supplementary Figure S1**. X-ray diffraction reciprocal space mapping for 20 nm $ErFeO_3$ thin film at (103) reflection

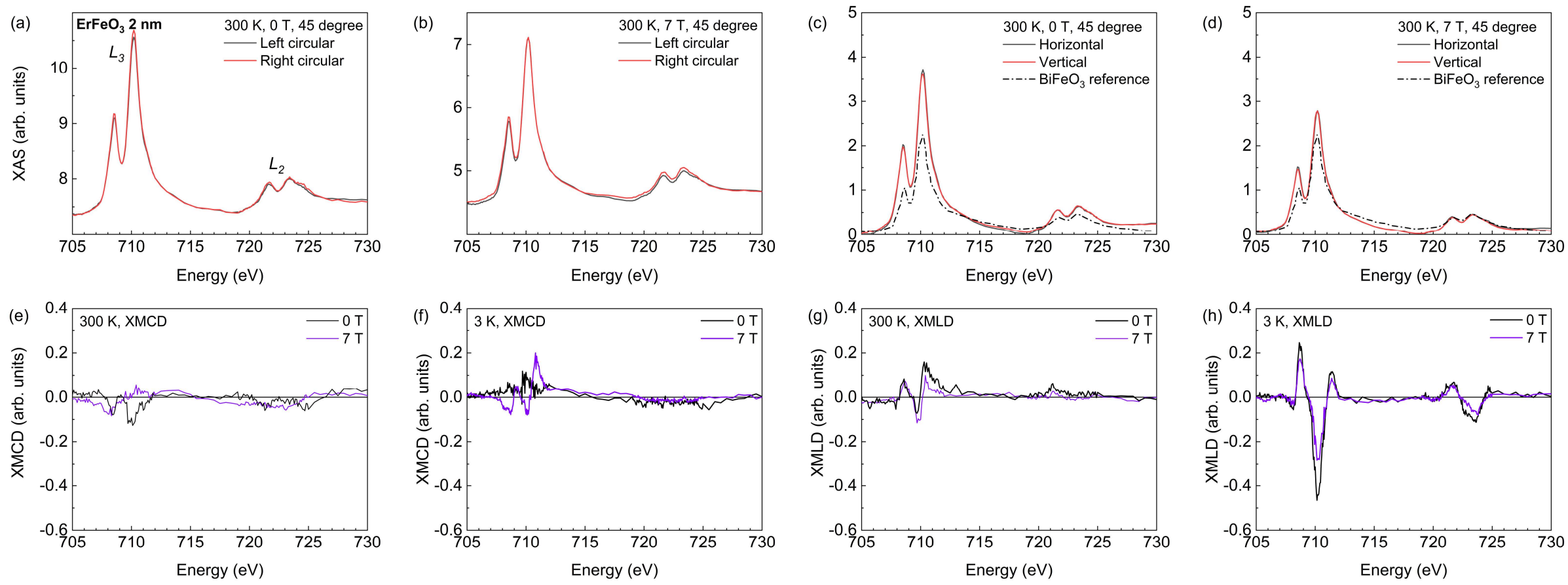

**Supplementary Figure S2**. X-ray absorption spectroscopy of 2-nm-thick $ErFeO_3$ film. (a,b) absorption spectra of left and right circular polarization at 300 K with 0, 7 T of magnetic field, respectively. (c,d) absorption spectra of horizontal and vertical linear polarization at 300 K with 0, 7 T of magnetic field respectively. (e,f) X-ray magnetic circular dichroism with 0, 7 T magnetic field at 300, 3 K, respectively. (g,h) X-ray magnetic linear dichroism with 0, 7 T magnetic field at 300, 3 K, respectively. The incident angle and magnetic field direction is 45°.

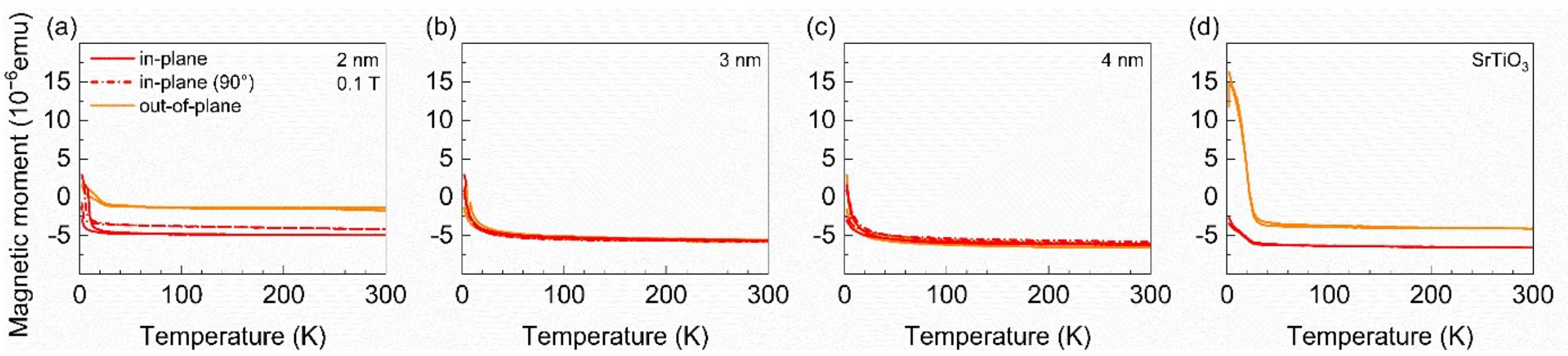

**Supplementary Figure S3**. Temperature dependent magnetization $M(T)$ data of (a) 2-, (b) 3-, and (c) 4-nm-thick $ErFeO_3$ films and (d) $SrTiO_3$ substrate with 0.1 T of magnetic field. Note that bare $SrTiO_3$ was measured to qualitatively assess the diamagnetic contribution rather than quantitatively compared to $ErFeO_3$ films since it was not exposed to thermal condition for $ErFeO_3$ film growth.

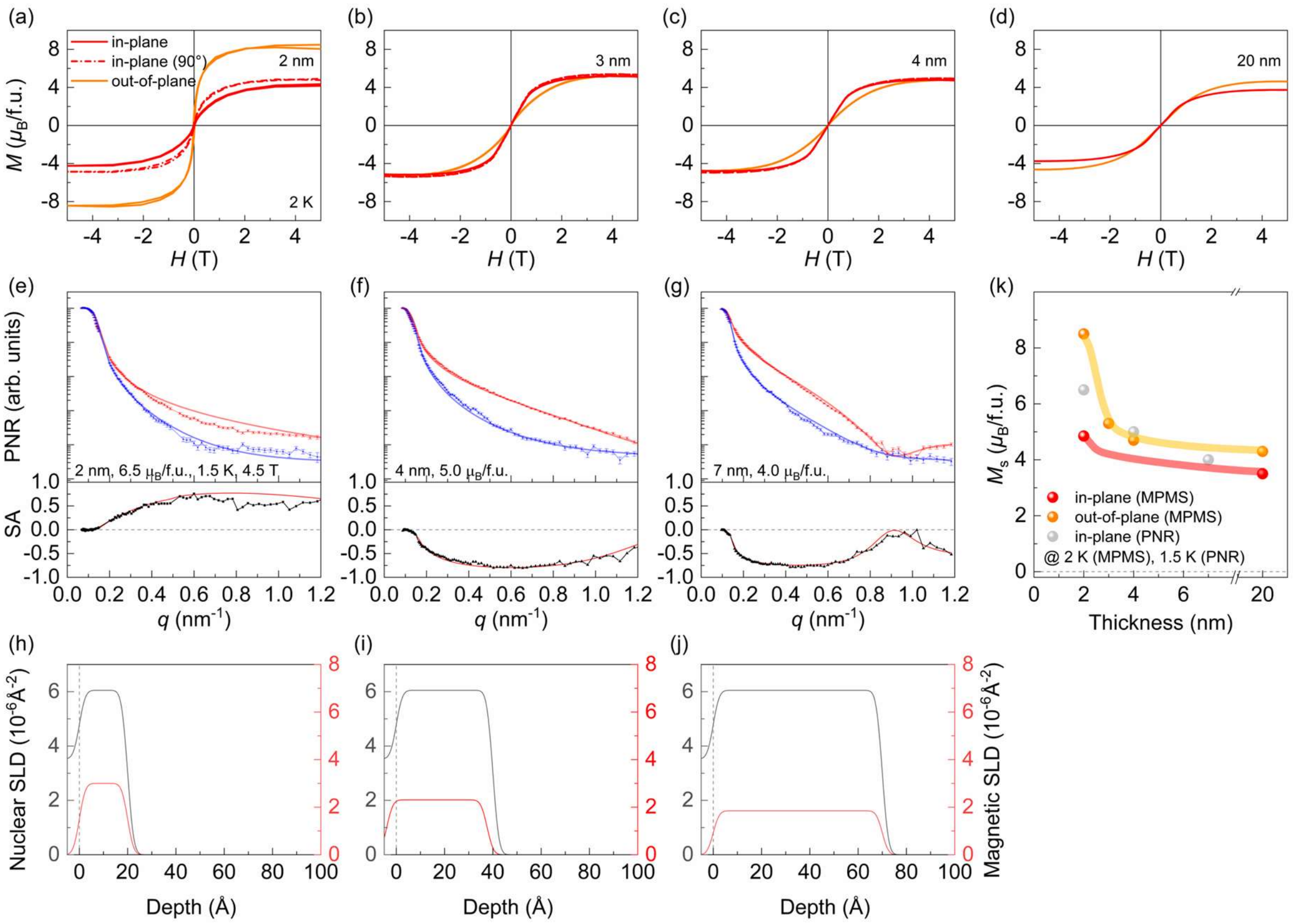

**Supplementary Figure S4**. (a-d) Magnetic property measurement system results of 2.0, 3.0, 4.0, 20 nm $ErFeO_3$ films at 2 K, (e-g) polarized neutron reflectivity results of 2, 4, 7 nm $ErFeO_3$ films at 1.5 K with 4.5 T, (h) Set of saturation magnetization value with respect to film thickness from magnetic property measurement system and polarized neutron reflectivity.

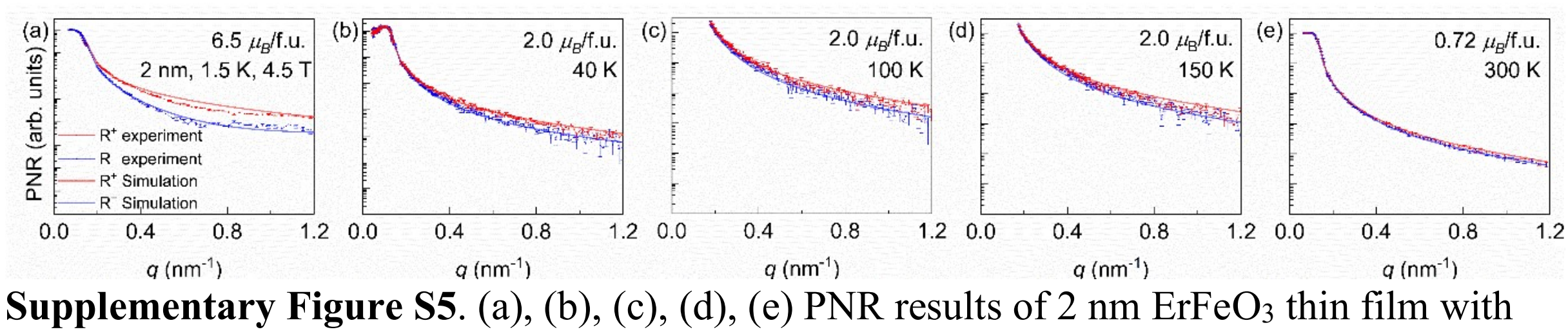

**Supplementary Figure S5**. (a), (b), (c), (d), (e) PNR results of 2 nm $ErFeO_3$ thin film with 4.5 T of magnetic field at 1.5, 40, 100, 150, 300 K, respectively.

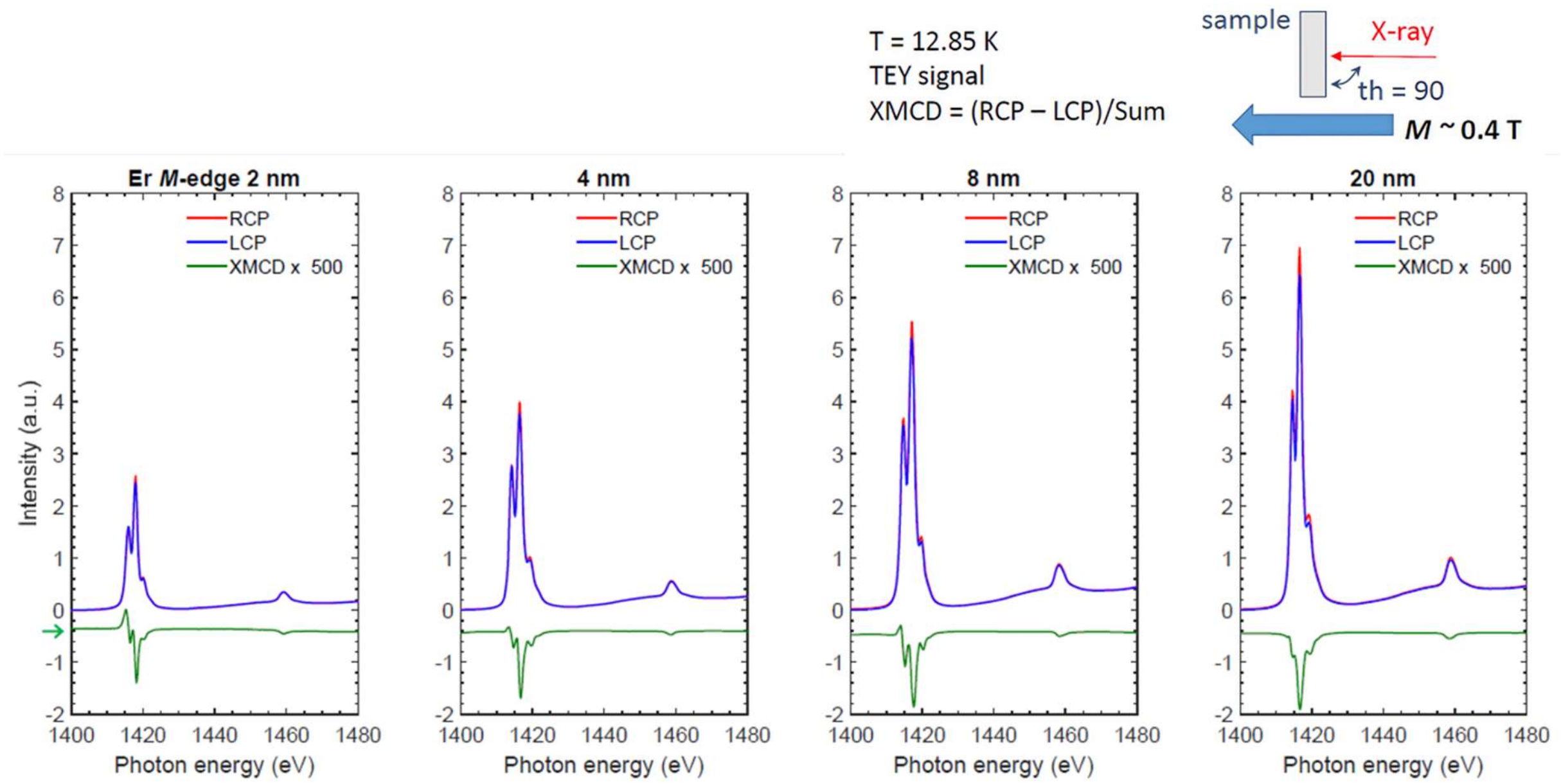

**Supplementary Figure S6**. X-ray magnetic circular dichroism of $ErFeO_3$ thin films with respect to film thickness at Er *M*-edge, 12.85 K. Experiment is conducted at Stanford Synchrotron Radiation Lightsource.

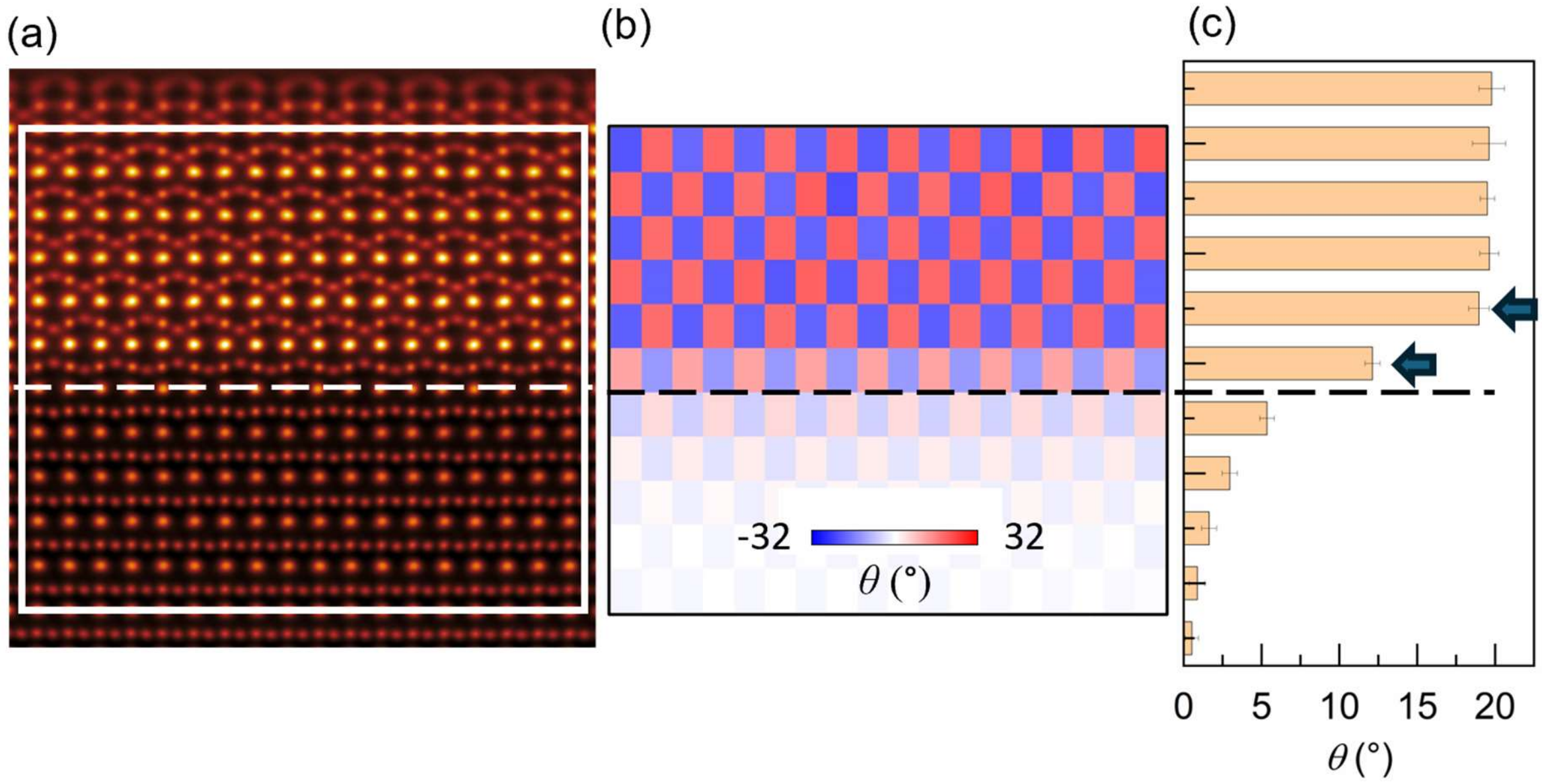

**Supplementary Figure S7**. (a) Scanning transmission electron microscopy image of ~2.3 nm of $ErFeO_3$ thin film, (b) Color mapping of $\theta$ (c) $\theta$ values of each layer.

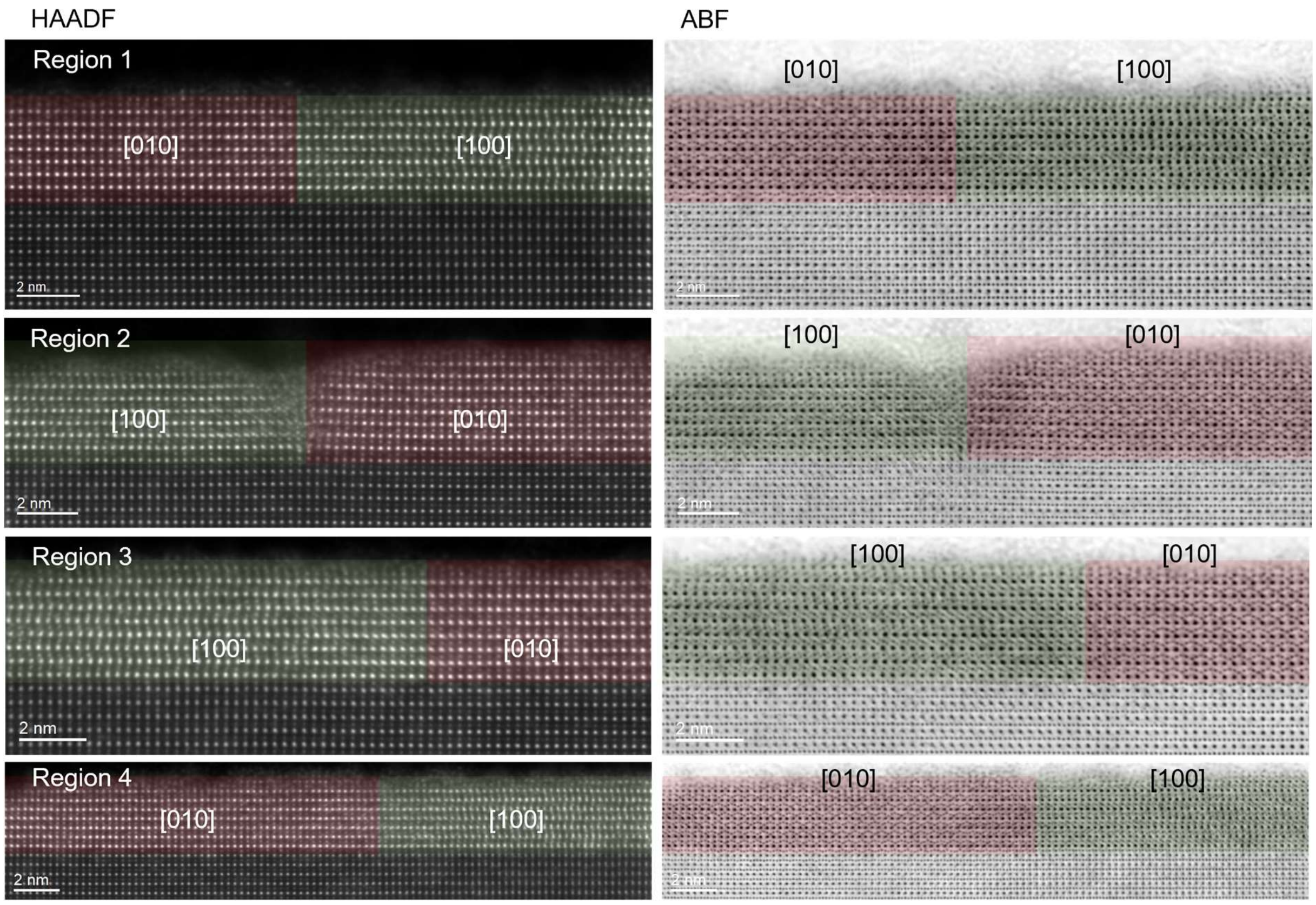

**Supplementary Figure S8**. High-angular annular dark-field scanning transmission electron microscopy, annular bright field image of ~2.3-nm-thick $ErFeO_3$ film for different regions

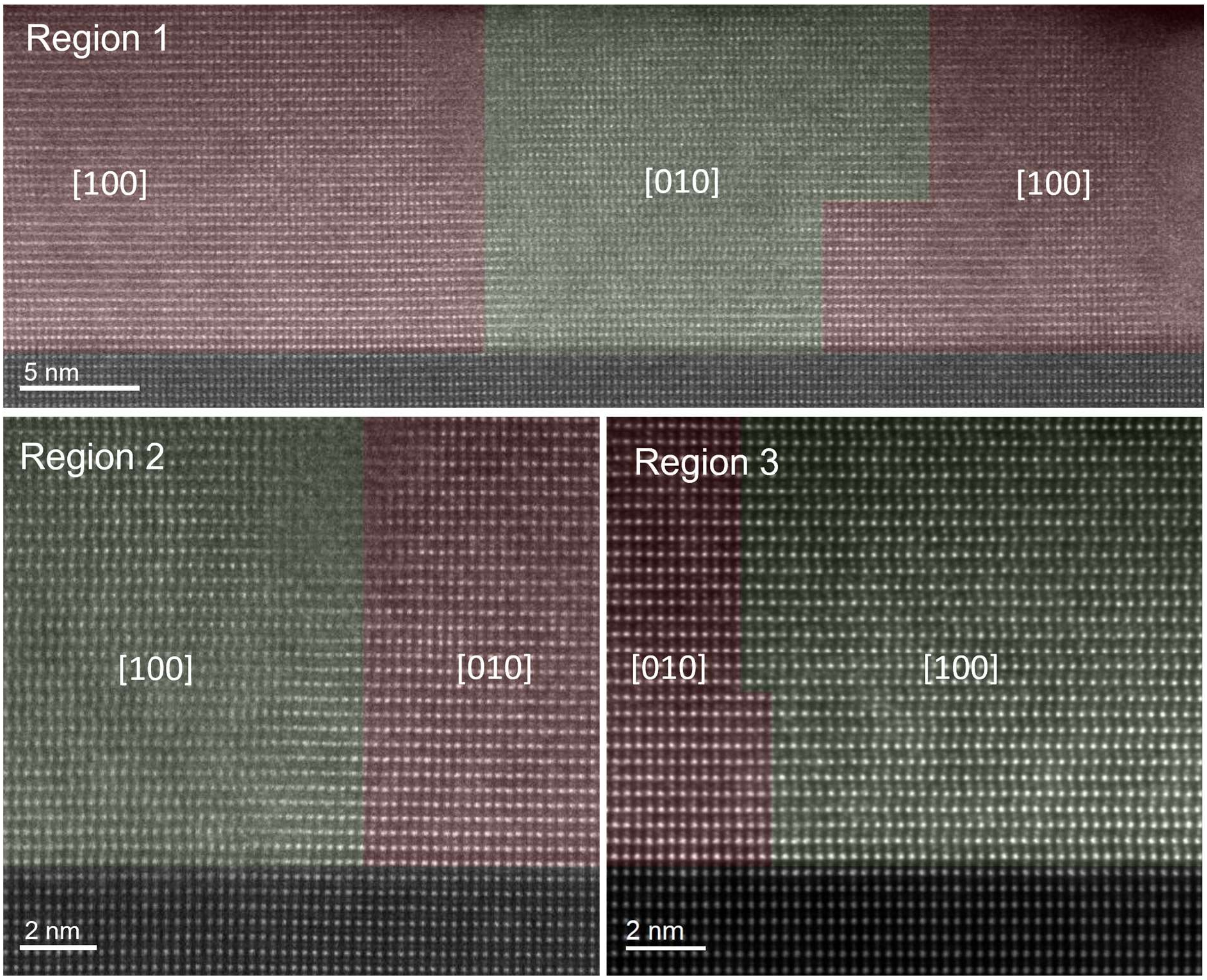

**Supplementary Figure S9**. High-angular annular dark-field scanning transmission electron microscopy image of ~13-nm-thick $ErFeO_3$ film.

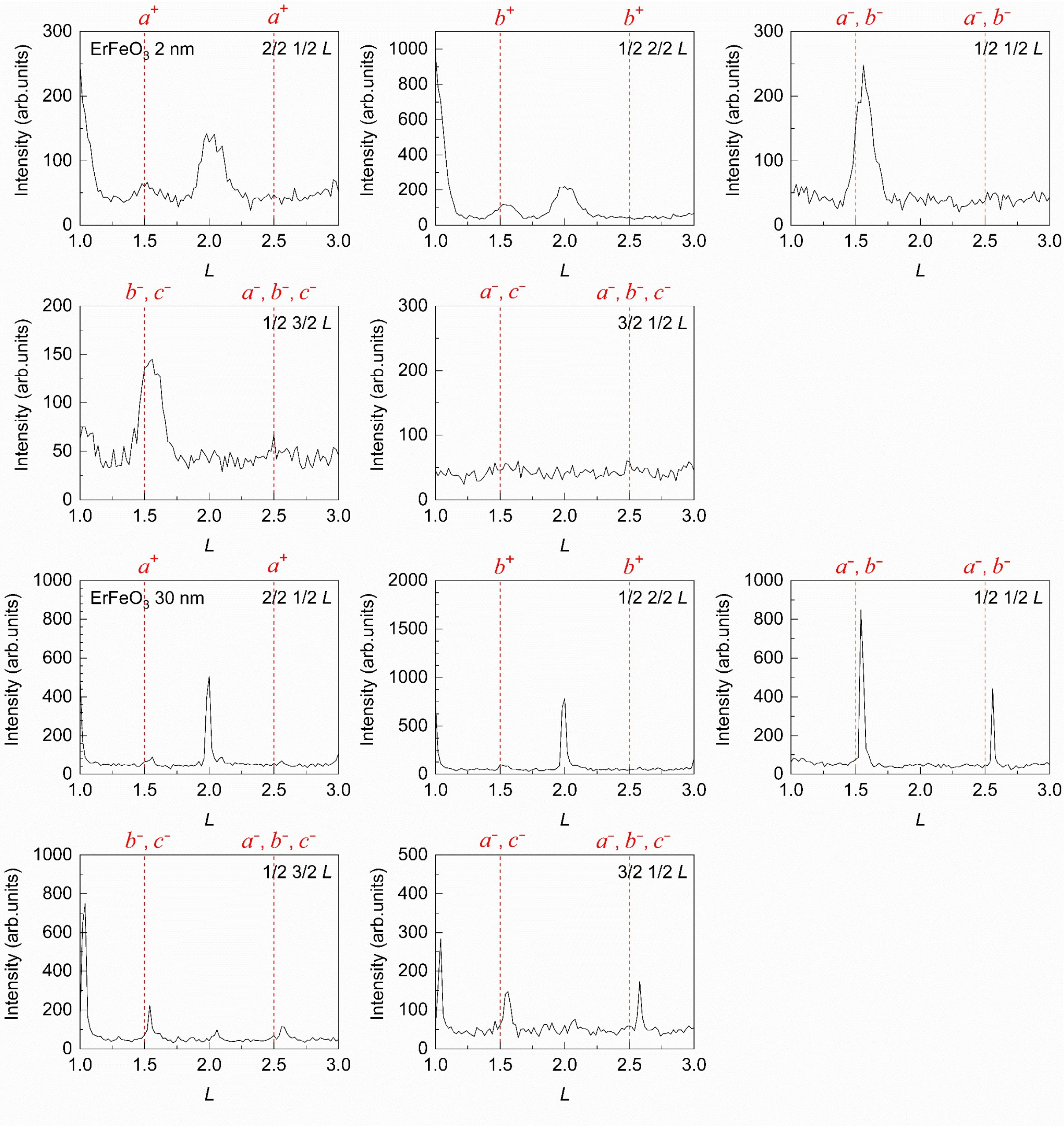

**Supplementary Figure S10**. X-ray diffraction half-order analyses on reciprocal space region of (2/2, 1/2, *L*), (1/2, 2/2, *L*), (1/2, 1/2, *L*), (1/2, 3/2, *L*), (3/2, 1/2, *L*) for $ErFeO_3$ 2 nm and 30 nm films.

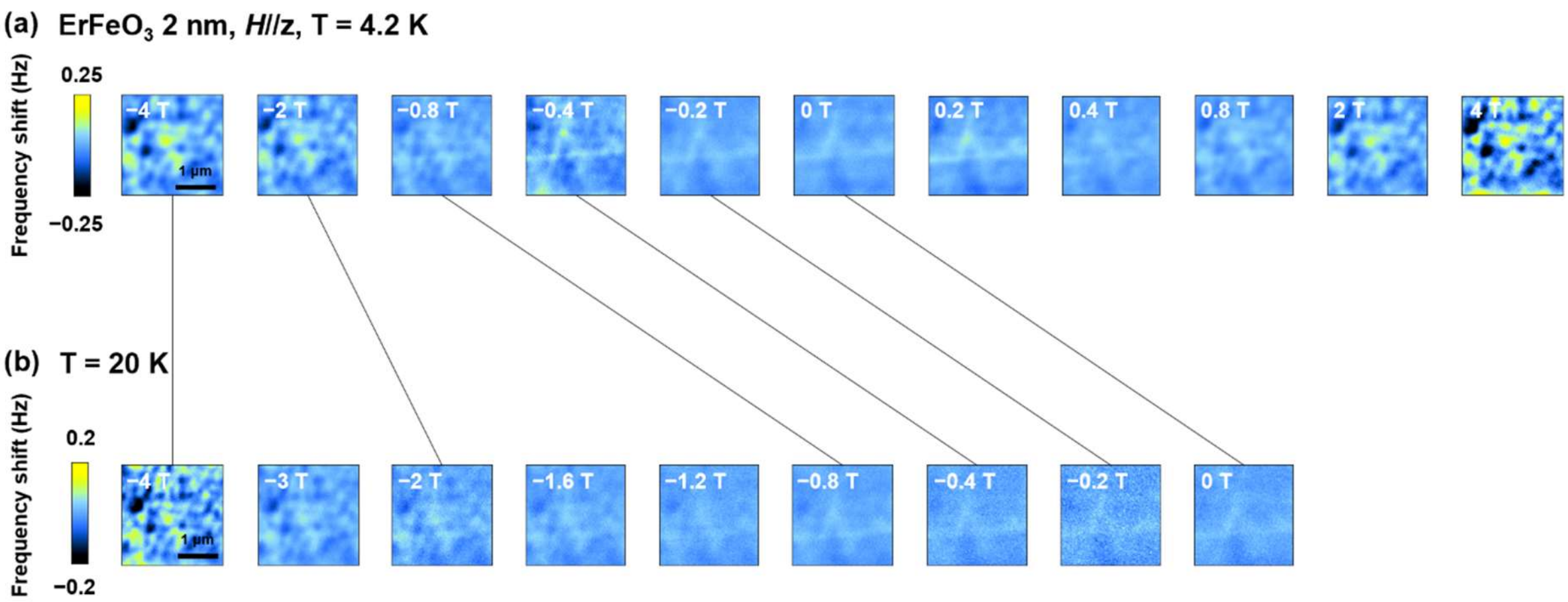

**Supplementary Figure S11.** (a), (b) Magnetic force microscopy images of $ErFeO_3$ 2-nm-thick film at 4.2 and 20 K. The same frequency-shift color scale was used for all images acquired at each temperature. Images obtained at the same magnetic field are connected by solid lines because the magnetic-field measurement points differed between the two temperatures.

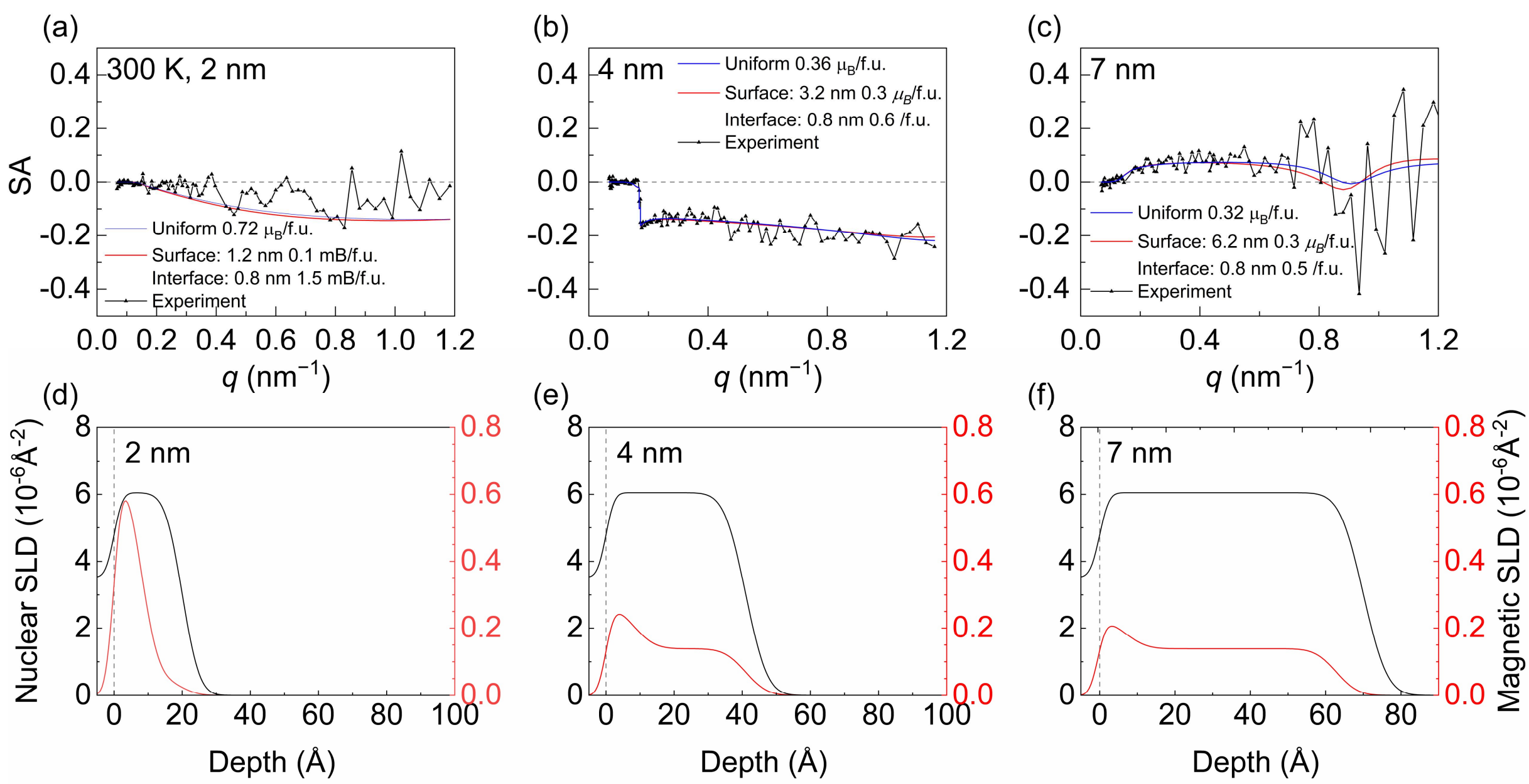

**Supplementary Figure S12**. (a), (b), (c) Polarized neutron reflectivity spin asymmetry plot of 2-, 4, and 7-nm-thick film at 300 K to compare uniform magnetic moment. (d), (e), (f) nuclear, and magnetic scattering length density of model with larger magnetic moment near the interface.

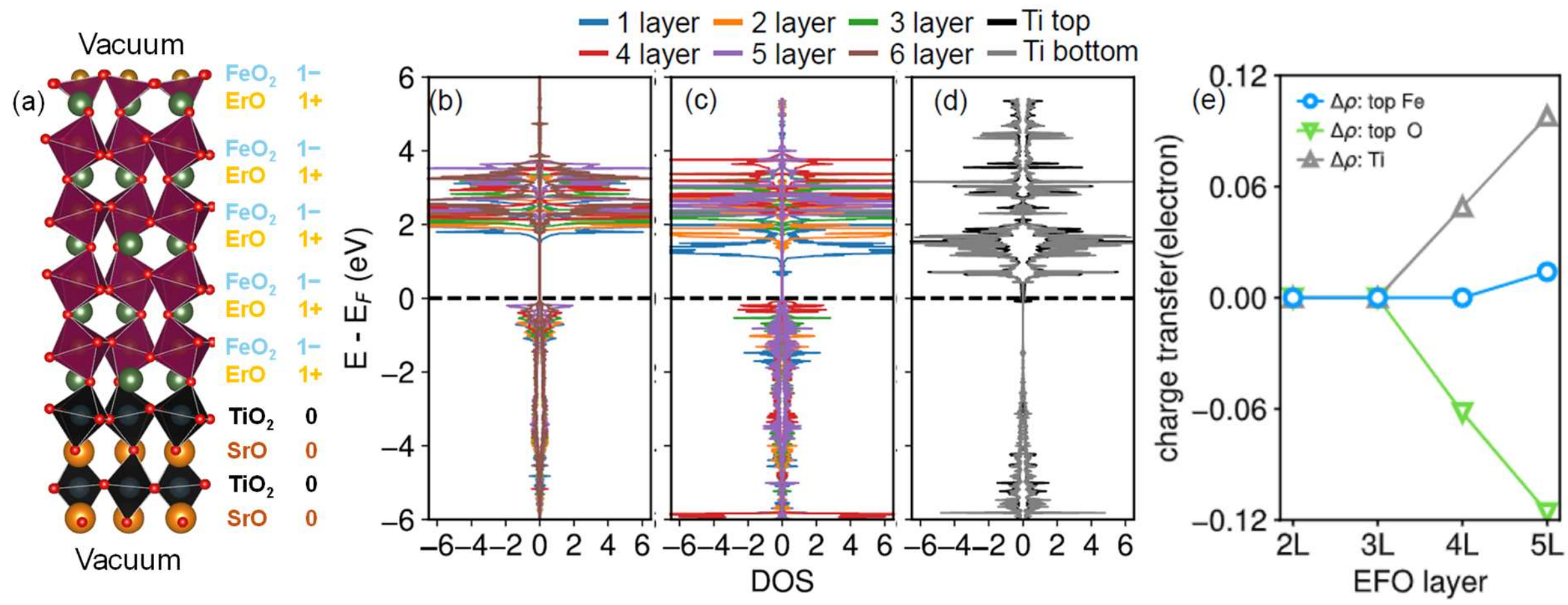

**Supplementary Figure S13.** (a) Schematic of the surface-terminated film with layer-by-layer formal charges, showing the alternating $(ErO)^{1+}/(FeO_2)^{1-}$ stack bounded by vacuum and the resulting potential gradient. (b,c) Layer-resolved Fe 3d-projected DOS of the film for $ErFeO_3$ thicknesses from 1 to 6 layers (color-coded by layer), revealing the thickness-dependent band shift associated with the polar potential gradient. (d) Ti 3d-projected DOS for the top and bottom Ti layers, showing the surface (top) Ti d-band approaching the Fermi level as the thickness increases. (e) Bader charge transfer Δρ for the topmost Fe, the topmost O, and Ti, plotted versus $ErFeO_3$ layer number.

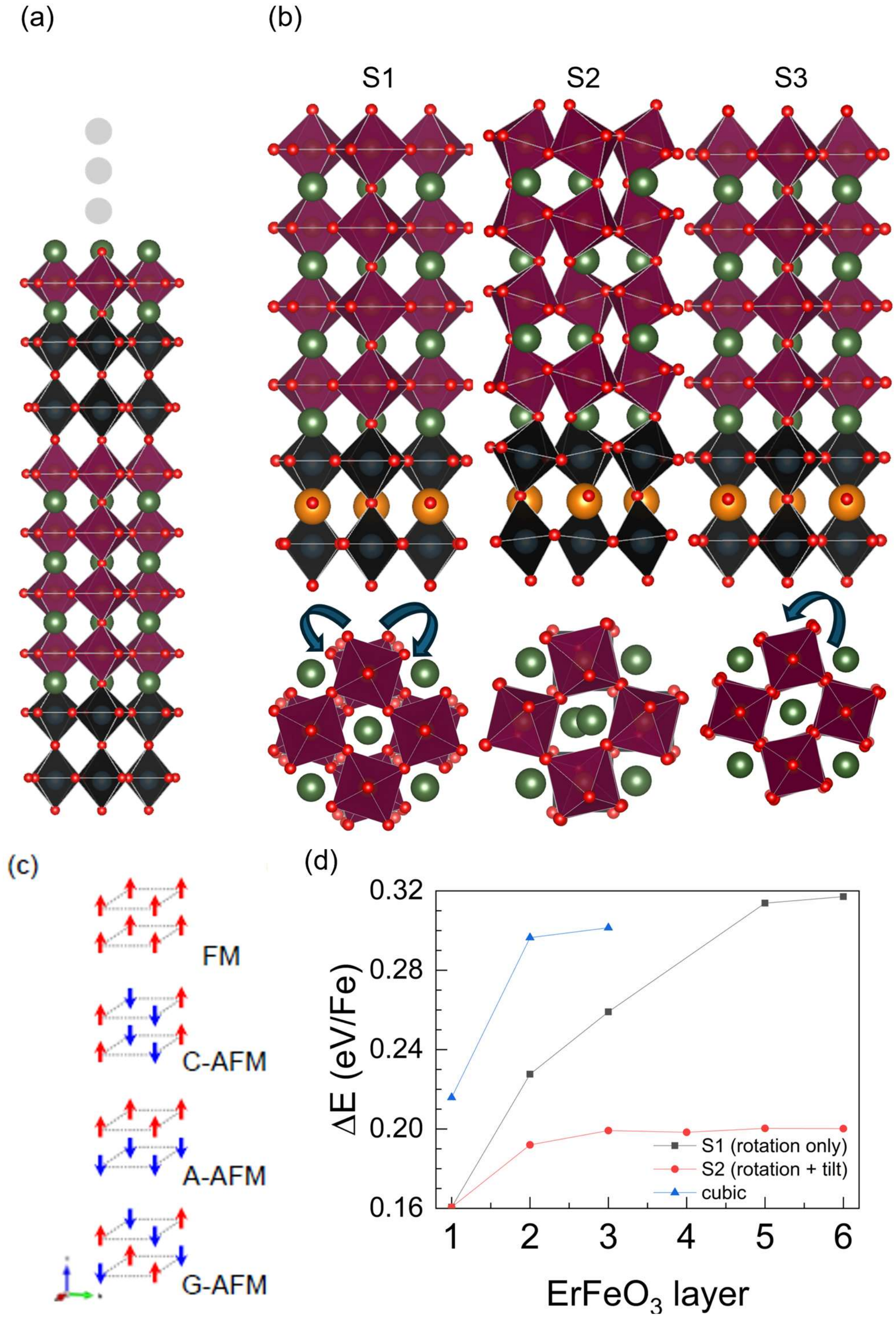

**Supplementary Figure S14.** Computational setup and layer-dependent magnetic stability of the periodic $[ErFeO_3]_n/[SrTiO_3]_2$ superlattice. (a) Atomic-structure schematic of the periodic superlattice stacking (shown for $n = 4$), with alternating $ErFeO_3$ and $SrTiO_3$ blocks. (b) The three $FeO_6$ rotation/tilt patterns considered —Str. 1 ($a^0a^0c^-$), bulk-like Str. 2 ($a^+b^-b^-$), and Str. 3 ($a^0a^0c^+$) — in side (top row) and top (bottom row) views. (c) The four collinear Fe-spin configurations sampled at every thickness: FM, C-AFM, A-AFM, and G-AFM. (d) Calculated $E_{FM} - E_{G\text{-}AFM}$ (in eV/Fe) versus $ErFeO_3$ layer number $n$ for three structural channels (rotation+tilt Str. 2, an additional rotation channel, and the non-rotation cubic $a^0a^0a^0$ reference), showing the progressive recovery of bulk-like antiferromagnetism with increasing thickness.

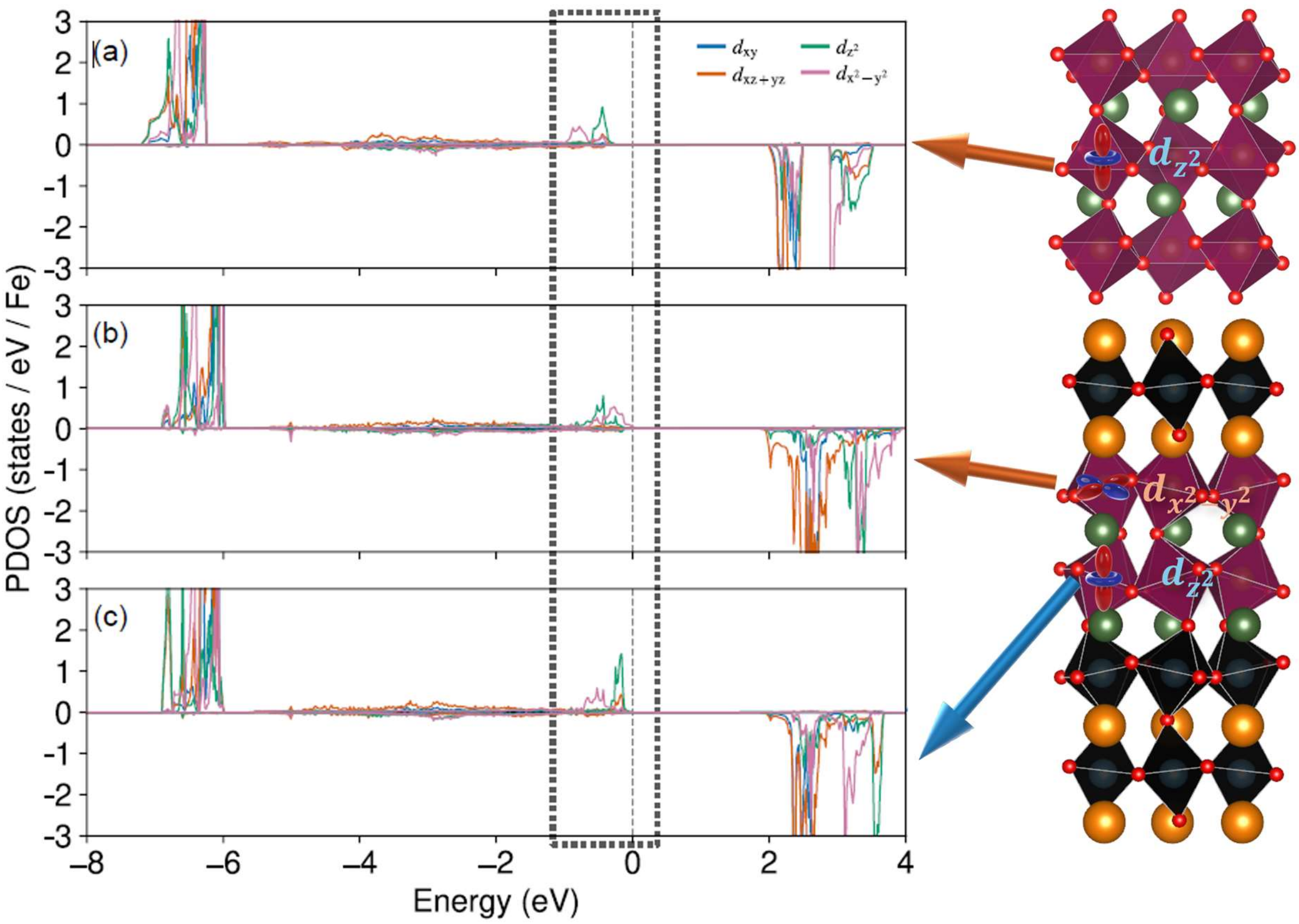

**Supplementary Figure S15.** Surface $e_g$-orbital reconstruction in the periodic $[ErFeO_3]_n/[SrTiO_3]_2$ superlattice. (a) Fe *d*-orbital-projected DOS for a bulk-like Fe, showing dominant $d_{z^2}$ character at the top of the valence band; (b) the same projection for the topmost Fe of the $ErFeO_3$ block, showing dominant $d_{x^2-y^2}$ character driven by the under-bonded $Sr^{2+}$ cation across the $ErFeO_3/SrTiO_3$ interface; (c) the corresponding projection for a subsurface Fe. The top and bottom structures on the right show schematic diagram of the bulk $ErFeO_3$ and $[ErFeO_3]_2/[SrTiO_3]_2$ superlattices, respectively. The schematic highlights the change of the orbital polarization of the top Fe in the $[ErFeO_3]_n/[SrTiO_3]_2$ film.

**Supplementary Note S1. Polar-discontinuity scenario in the $ErFeO_3/SrTiO_3$ film geometry**

In addition to the surface $e_g$-orbital reconstruction discussed in the main text, and here we examine the electronic and magnetic consequences of the polar discontinuity in the $ErFeO_3/SrTiO_3$ film. The film geometry consists of an SrO-terminated $SrTiO_3$ side and an $FeO_2$-terminated vacuum side, with the $ErFeO_3$ thickness varied from n = 1 to 6 (Figure S13a). The relaxed $FeO_6$ rotation/tilt patterns are essentially identical between the film and the periodic superlattice (Supplementary Note S2) for n ≥ 2, so the differences reported below originate from the surface/interface electronic environment rather than from structural properties.

Along [001] direction the film consists of alternating $(ErO)^{1+}$ and $(FeO_2)^{1-}$ atomic planes, which makes it electrically polar in close analogy to the well-known $LaAlO_3/SrTiO_3$ heterostructure. The layer-resolved Fe and Ti d-projected densities of states (Figure S13(b–d)) reveal a pronounced potential gradient across the film, and around n = 3, the top Fe and bottom Ti bands start to touch the Fermi level, signaling an incipient surface metallicity that compensates the polar field. A Bader-charge analysis (Figure S13e) shows that the charge transfer mostly occurs between the top-surface oxygen and the bottom Ti, leaving the topmost Fe charge status essentially unchanged. This follows naturally from the charge-transfer-insulator character of $ErFeO_3$, whose valence-band edge is dominated by O 2p states with only minor Fe 3d weight, so that the compensating holes reside mainly on the top planar oxygen. This is consistent with the Fe *L*-edge XAS, which detects only $Fe^{3+}$ in the ultrathin films.

The polar field also leaves a distinct character in the energetics between the ferromagnetic and G-AFM state. In the rotation + tilt (Str. 2, $a^+b^-b^-$) film geometry, the calculated $E_{FM} - E_{G\text{-}AFM}$ is non-monotonic in thickness: it decreases from 0.22 eV/Fe at n = 2 to a minimum of 0.16 eV/Fe near n = 4, and then rises again toward 0.23 eV/Fe at n = 6 (Figure 4f). This reflects competition between two opposing contributions as the film becomes thicker: additional bulk-like Fe layers favor G-type antiferromagnetism, whereas the growing band bending due to the polar discontinuity increases the spectral weight of the surface metallic band at the Fermi level and hence the ferromagnetism-promoting itinerant exchange. The crossover near n ~ 4–5 marks the thickness at which the bulk-like antiferromagnetic contribution overcomes the surface-metallic ferromagnetic one. In the opposite (thinnest) limit, on the other hand, the polar band bending reduces with decreasing thickness, so that below n ~ 4 this scenario predicts a recovery of the antiferromagnetic stability rather than its suppression. Hence the polar-discontinuity mechanism would tend to re-stabilize antiferromagnetism in the thinnest films, somewhat consistent with the reduction of $M_s$ at 1.2 nm of the film thickness (see Fig. 2(a) and (k) in the manuscript). We stress that no signature of the surface metallicity required by this scenario has been resolved in the present experiments; the polar-discontinuity picture is therefore retained only as a theoretical possibility rather than a confirmed mechanism.

## Supplementary Note S2. Periodic $ErFeO_3/SrTiO_3$ superlattice calculations

As a complementary reference to the surface-terminated film, we also carried out calculations on a periodic $ErFeO_3/SrTiO_3$ superlattice, in which $[ErFeO_3]_n$ blocks (n = 1–6) alternate with two-unit-cell $[SrTiO_3]_2$ blocks without any vacuum surface or polar discontinuity (Figure S14a). As in the film, the in-plane lattice constant was fixed to the cubic $SrTiO_3$ value ($a_{STO}$ = 3.905 Å) to enforce the experimental epitaxial strain. At every thickness we sampled four collinear Fe-spin configurations (FM, A-AFM, C-AFM, and G-AFM) in combination with three $FeO_6$ rotation/tilt patterns —Str. 1 ($a^0a^0c^-$), bulk-like Str. 2 ($a^+b^-b^-$), and Str. 3 ($a^0a^0c^+$), — to locate the structural and magnetic ground state (Figure S14b and S14c). The Fe magnetic moment is robust at 4.16 $\mu_B$/Fe in the G-AFM ground state and 4.29 $\mu_B$/Fe in the metastable FM state, consistent with high-spin $Fe^{3+}$ ($d^5$, S = 5/2). The G-AFM ordering combined with the bulk-like $a^+b^-b^-$ tilt pattern (Str. 2) is the ground state at every thickness. As shown in Figure S14d, $E_{FM} - E_{G\text{-}AFM}$ in the Str. 2 channel grows from 0.16 eV/Fe at n = 1 to 0.20 eV/Fe at n = 6, indicating that the bulk-like antiferromagnetic superexchange is gradually recovered as more $ErFeO_3$ unit cells are stacked. When all $FeO_6$ rotations and tilts are switched off (the cubic $a^0a^0a^0$ reference), the separation increases substantially, reaching 0.30 eV/Fe at n = 2–3, while the rotation-only Str. 1 pattern lies between the cubic and the rotation + tilt cases. This ordering follows the Goodenough–Kanamori rule: the cubic 180° Fe–O–Fe geometry maximizes the σ-bond superexchange overlap and yields the strongest antiferromagnetism, whereas the orthorhombic $a^-a^-c^+$ tilt bends the bond and weakens the antiferromagnetic channel. The superlattice also reproduces the surface $e_g$-orbital reconstruction emphasized in the main text. The topmost Fe of the $ErFeO_3$ block exhibits a clear $d_{x^2-y^2}$ character at the top of the valence band, in place of the bulk-like $d_{z^2}$ character of the deeper Fe layers (Figure S14). In the superlattice this orbital order is driven by the $Sr^{2+}$ cation across the $ErFeO_3/SrTiO_3$ interface, in direct analogy to the missing apical oxygen that produces the same reconstruction at the vacuum surface of the film. The resulting vertical Fe–Fe superexchange again proceeds between orthogonal $d_{x^2-y^2}$ and $d_{z^2}$ orbitals, which by the Goodenough–Kanamori rules is only weakly antiferromagnetic or even ferromagnetic. Finally, we note that in both the superlattice and the film calculations structural relaxation favors the bulk-like rotation + tilt (Str. 2) pattern at every thickness, in apparent contradiction with the experimental observation that the $FeO_6$ tilt is suppressed in the thinnest layers and develops only above 2 nm. This likely reflects an overestimation of the structural rigidity in the static, zero-temperature calculations, and is a further reason why the present first-principles results are used to motivate qualitative mechanisms rather than to provide a quantitative account of the thickness dependence.